\pdfoutput=1
\documentclass[%
aps,
superscriptaddress,
two column,
reprint,%
]{revtex4-1}

\usepackage{graphicx}% Include figure files
\usepackage{dcolumn}% Align table columns on decimal point
\usepackage{bm}% bold math
\usepackage{epstopdf}
\usepackage{adjustbox}

\usepackage[utf8]{inputenc}
\usepackage[T1]{fontenc}
\usepackage{mathptmx}
\usepackage{upgreek}
\usepackage{textcomp}
\usepackage{gensymb}
\usepackage{anyfontsize}
\usepackage{amsmath}
\usepackage{hyperref}
\usepackage{makecell}

\begin{document}

\title{Coplanar cavity for strong coupling between photons and magnons in van der Waals antiferromagnet}
\author{Supriya Mandal}
\thanks{these authors contributed equally}
\affiliation{Department of Condensed Matter Physics and Materials Science, Tata Institute of Fundamental Research, Homi Bhabha Road, Mumbai 400005, India}
\author{Lucky N. Kapoor}
\thanks{these authors contributed equally}
\affiliation{Department of Condensed Matter Physics and Materials Science, Tata Institute of Fundamental Research, Homi Bhabha Road, Mumbai 400005, India}
\author{Sanat Ghosh}
\affiliation{Department of Condensed Matter Physics and Materials Science, Tata Institute of Fundamental Research, Homi Bhabha Road, Mumbai 400005, India}
\author{John Jesudasan}
\affiliation{Department of Condensed Matter Physics and Materials Science, Tata Institute of Fundamental Research, Homi Bhabha Road, Mumbai 400005, India}
\author{Soham Manni}
\affiliation{Department of Condensed Matter Physics and Materials Science, Tata Institute of Fundamental Research, Homi Bhabha Road, Mumbai 400005, India}
\affiliation{Department of Physics, Indian Institute of Technology Palakkad, Palakkad 678557, India}
\author{A. Thamizhavel}
\affiliation{Department of Condensed Matter Physics and Materials Science, Tata Institute of Fundamental Research, Homi Bhabha Road, Mumbai 400005, India}
\author{Pratap Raychaudhuri}
\affiliation{Department of Condensed Matter Physics and Materials Science, Tata Institute of Fundamental Research, Homi Bhabha Road, Mumbai 400005, India}
\author{Vibhor Singh}
\thanks{Corresponding author}
\email{v.singh@iisc.ac.in}
\affiliation{Department of Physics, Indian Institute of Science, Bangalore 560012, India}
\author{Mandar M. Deshmukh}
\thanks{Corresponding author}
\email{deshmukh@tifr.res.in}
\affiliation{Department of Condensed Matter Physics and Materials Science, Tata Institute of Fundamental Research, Homi Bhabha Road, Mumbai 400005, India}

\begin{abstract}
We investigate the performance of niobium nitride superconducting coplanar waveguide resonators towards hybrid quantum devices with magnon-photon coupling. We find internal quality factors $\sim$ 20000 at 20~mK base temperature, in zero magnetic field. We find that by reducing film thickness below 100~nm internal quality factor greater than 1000 can be maintained up to parallel magnetic field of $\sim$ 1~T and perpendicular magnetic field of $\sim$ 100~mT. We further demonstrate strong coupling of microwave photons in these resonators, with magnons in chromium trichloride, a van der Waals antiferromagnet, which shows that these cavities serve as a good platform for studying magnon-photon coupling in 2D magnonics based hybrid quantum systems. We demonstrate strong magnon-photon coupling for both optical and acoustic magnon modes of an antiferromagnet.
\end{abstract}
\maketitle

\renewcommand{\thesection}{\Roman{section}}
\setcounter{section}{0}

\renewcommand{\thefigure}{\arabic{figure}}
\setcounter{figure}{0}

\renewcommand{\theequation}{\arabic{equation}}
\setcounter{equation}{0}

\renewcommand{\thetable}{\Roman{table}}
\setcounter{table}{0}

Superconducting coplanar waveguide (SCPW) resonators are of interest as elements of architecture for hybrid quantum systems and quantum computation related studies. The two dimensional (2D) structure, scalability and control over impedance across varying length scales for SCPW provide a natural way to couple them with various mesoscopic systems. They have been used for making kinetic inductance detectors, \cite{day_broadband_2003} parametric amplifiers, \cite{tholen_nonlinearities_2007,castellanos-beltran_widely_2007} and have been coupled to superconducting qubits, \cite{wallraff_strong_2004} nano-mechanical resonators, \cite{regal_measuring_2008,singh_optomechanical_2014} spin ensembles, \cite{kubo_strong_2010,amsuss_cavity_2011,ranjan_probing_2013,tkalcec_strong_2014,zollitsch_high_2015} and quantum dots. \cite{petta_coherent_2005,nowack_coherent_2007} They have become an important component for realizing hybrid quantum devices.

For application of SCPW resonators in hybrid quantum devices involving electron spin resonance (ESR) systems, \cite{schuster_high-cooperativity_2010,ranjan_probing_2013,malissa_superconducting_2013,benningshof_superconducting_2013} nitrogen vacancy (NV) centers, \cite{kubo_strong_2010,amsuss_cavity_2011} and different topological systems, \cite{schmidt_ballistic_2018,kroll_magnetic_2018,wang_coherent_2019} performance under magnetic field is additionally required. Furthermore, with the advent of 2D magnetic materials \cite{gong_two-dimensional_2019} and possibility to utilize their unique properties in hybrid quantum designing schemes $\sim$ 100~mT  magnetic fields in arbitrary directions is required. This additional requirement for the SCPW resonators necessitates use of type-II superconductors having high upper critical magnetic field ($H_{C2}$) like Mo-Re, TiN, Nb-Ti-N and NbN. \cite{singh_molybdenum-rhenium_2014,calado_ballistic_2015,van_woerkom_one_2015,vissers_low_2010,carter_low-loss_2019} In the presence of magnetic field, flux vortices are generated in these materials which cause dissipative vortex motion in presence of high frequency oscillatory currents. Moreover, reduction in Cooper pairs due to formation of normal cores of the vortices causes increase in kinetic inductance which causes a decrease in resonance frequency. \cite{song_microwave_2009} Control over this kind of vortex induced dissipation can be achieved through introduction of pinning sites to trap the vortices which can either be done by utilizing intrinsic disorder \cite{ghirri_yba2cu3o7_2015} or by fabricating artificial pinning sites. \cite{song_reducing_2009,bothner_improving_2011,kroll_magnetic-field-resilient_2019}

\begin{figure}
\includegraphics{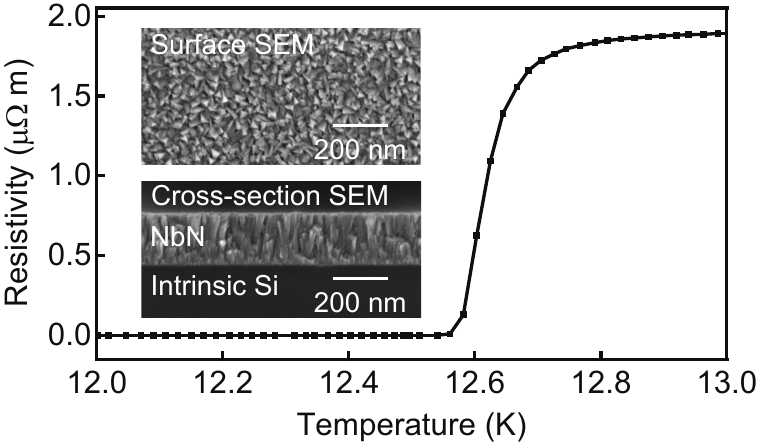}
\caption{\label{fig:figure1} Variation of resistivity of a 80 nm thick NbN film deposited by sputtering on intrinsic silicon with temperature [Inset: SEM images of top and cross sectional views of an NbN film].}
\end{figure}

Niobium nitride (NbN) is a disordered type-II superconductor with a high transition temperature ($T_C\approx16.8\ \textrm{K}$), high upper critical magnetic field ($H_{C2}\approx15\ \textrm{T}$), small coherence length ($\approx5\ \textrm{nm}$) and large penetration depth ($\approx250\ \textrm{nm}$). \cite{chockalingam_superconducting_2008,mondal_phase_2011,kamlapure_measurement_2010} NbN thin films of varying thicknesses with moderately high $T_C$ can be realized, which are also stable in ambient condition and robust with respect to thermal cycling from room temperature to cryogenic temperatures. Another interesting property of NbN is that its $T_C$ depends mainly on the carrier density (and not disorder) for samples with $T_C$ > 10 K. \cite{chockalingam_superconducting_2008} This makes NbN a potential candidate for reducing dissipation due to vortices by utilizing inherent disorder.

In this work we fabricate and probe the properties of NbN SCPW resonators. Traditionally, SCPW resonators are fabricated by careful substrate surface preparation followed by the deposition of superconducting film. Subsequently, etching processes are used to pattern the resonator. Here we intentionally use a simple lift-off based fabrication process. The motivation behind this approach is two-fold: (a) it helps to establish baseline properties and (b) it opens up the possibility to integrate exfoliable  materials into microwave circuitry. For these resonators we observe internal $Q$, $Q_i$ of $\sim 2\times 10^4$ at 20 mK temperature and zero magnetic field. We study the magnetic field dependence of quality factor and resonance frequency dispersion of these resonators, primarily for two different thicknesses $(t)$ of the NbN thin film. We find that $Q_i$ is higher than 1000 for in-plane fields of $\sim1$~T and perpendicular fields of $\sim 100$~mT. Few 100 mT is often the magnetic field range around which many materials show ferromagnetic, antiferromagnetic and electron spin resonances. \cite{macneill_gigahertz_2019,zhang_strongly_2014,kapoor_observation_2021} For example, using average $g$-factor of 2.2 for chromium trihalides \cite{kim_evolution_2019} we find that approximately 160 mT magnetic field is required for resonant coupling of the magnons in these materials with microwave photons in a resonator with resonance frequency around 5 GHz. This makes NbN SCPW resonators an optimal platform for studying collective spin oscillations in different materials. Here we demonstrate coupling between a cavity mode in an NbN SCPW resonator and antiferromagnetic resonance (AFMR) modes in a chromium trichloride (CrCl$_3$) crystal -- a van der Waals antiferromagnet. \cite{macneill_gigahertz_2019,kapoor_observation_2021}

We use intrinsic silicon as substrate for fabrication of the SCPW resonators (we also fabricated resonators on sapphire substrate and the corresponding characterization data is included in the supplementary material). First, the wafers are spin-coated with a bilayer resist consisting of EL9 and PMMA 950 A2. Subsequently, resonator pattern is made by electron beam lithography and developing in a solution of MIBK and IPA in 1:3 volume ratio. NbN is then deposited by reactive DC sputtering with a niobium (Nb) target at a sputtering power of 230 W in presence of continuous flow of 11 sccm of nitrogen and 70 sccm of argon at a sputtering pressure of $6.5\times10^{-3}\ \textrm{mbar}$ and temperature $120\ \degree\textrm{C}$. \cite{chockalingam_superconducting_2008} After sputtering, lift-off is done in acetone to remove the remaining resist along with excess NbN film on top of them. On an NbN film of thickness 80 nm, transport measurements give a $T_C\approx12.5 \textrm{ K}$, $H_C>14 \textrm{ T}$, $\textrm{RRR}\approx0.9$ and room temperature resistivity $\approx1.95\ \upmu\upOmega\ \textrm{m}$. Variation of resistivity with temperature for the NbN film on intrinsic silicon substrate is shown in Fig.~\ref{fig:figure1}. Fig.~\ref{fig:figure1} inset shows SEM image of an NbN film where disorder is clearly visible.

\begin{figure}[h]
\includegraphics{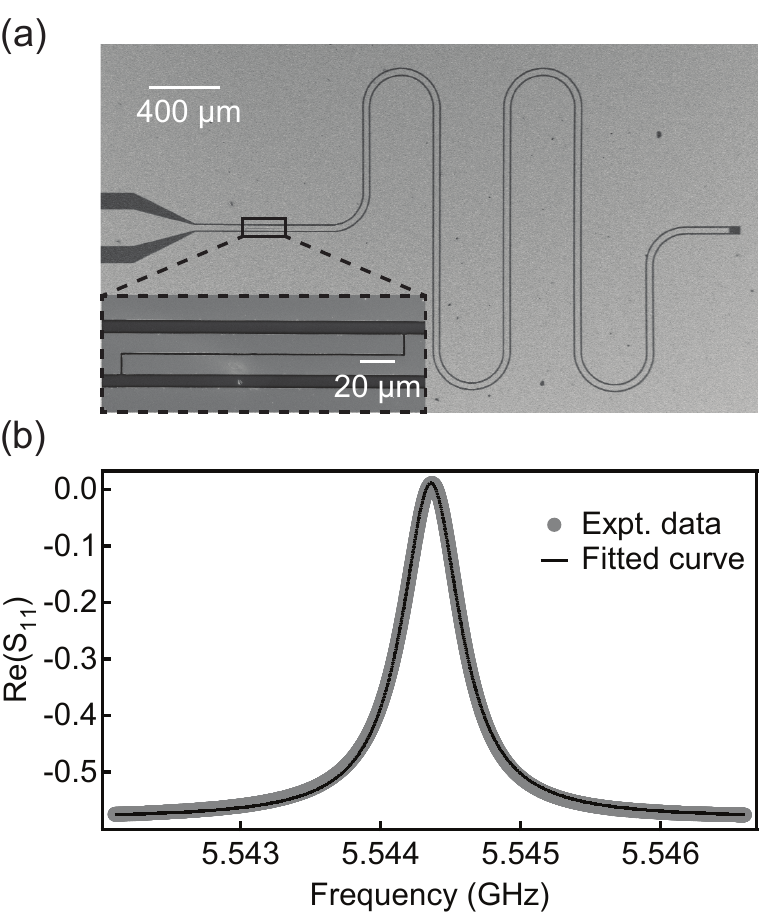}
\caption{\label{fig:figure2} (a) SEM image of an NbN SCPW resonator [Inset: Zoomed in image of coupling capacitor]. (b) Real part of $S_{11}$ measured at $20\ \textrm{mK}$ temperature along with fitting function.}
\end{figure}

The SCPW resonators are designed in half-wave single port configuration. Fig.~\ref{fig:figure2} (a) shows SEM image of the device. The resonator has a trace width of 28 $\upmu$m and has been designed from a section of coplanar waveguide with characteristic impedance of 46 $\upOmega$ on intrinsic silicon, neglecting the surface impedance of the superconducting film. A coupling capacitor is made between input port and the resonator for coupling microwave power in-and-out of the resonator. The measurements are done in a dilution fridge and sufficient number of attenuators are kept at different plates of the fridge in the input line to ensure proper thermalization of microwave photons reaching the sample (more details in the supplementary material).

\begin{figure*}
\includegraphics{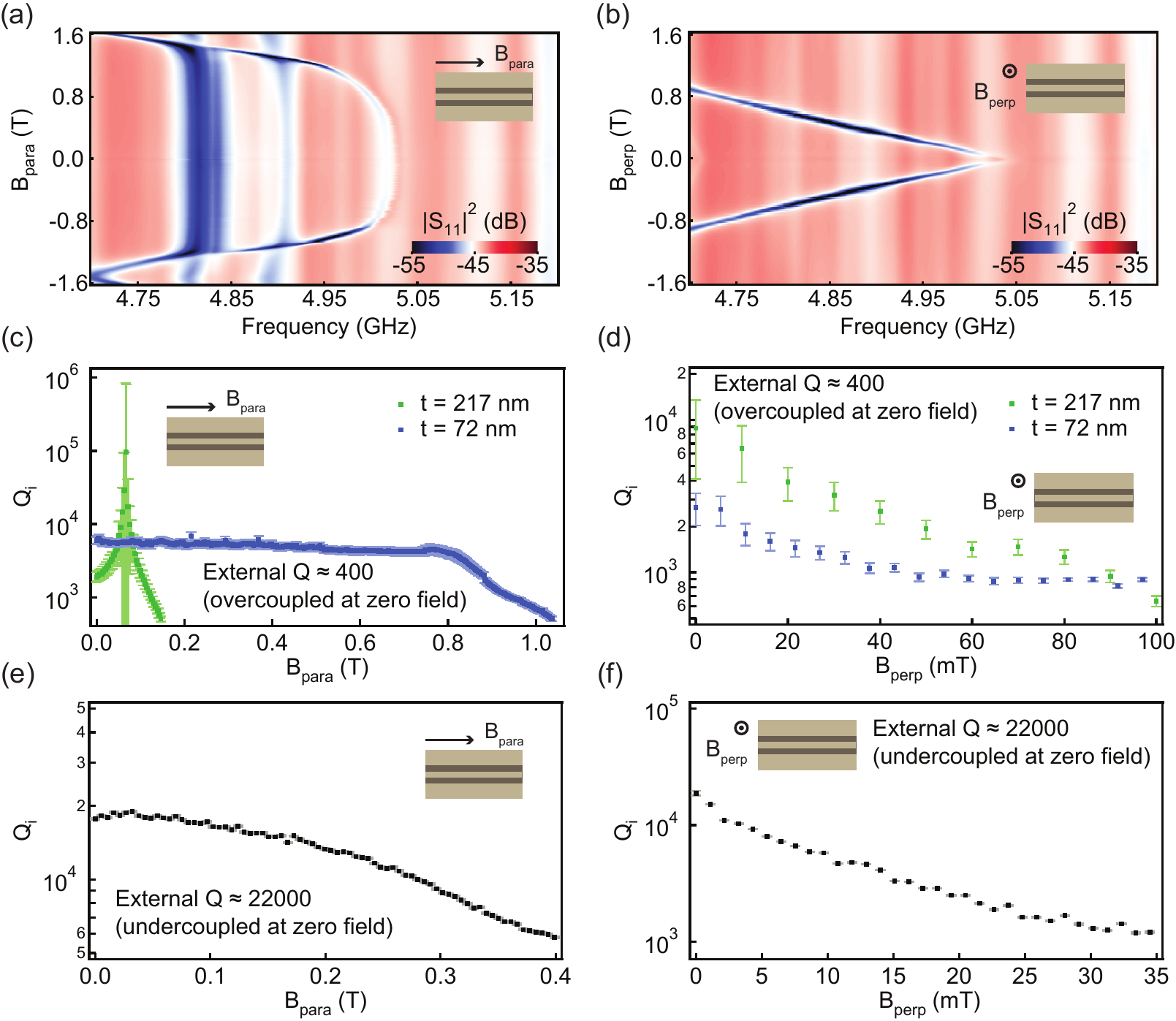}
\caption{\label{fig:figure3} Characterization of NbN resonators as a function of magnetic field. (a) and (b) Color-scale plots showing variation in $\lvert S_{11} \rvert^2$ for an NbN SCPW resonator with NbN film thickness $72\ \textrm{nm}$, with frequency and magnetic field for field orientations parallel and perpendicular to the SCPW plane respectively. (c) and (d) Variation of $Q_i$ with magnetic field for film thicknesses 217 nm and 72 nm on intrinsic silicon substrate, for parallel and perpendicular field orientations respectively. These resonators are overcoupled, $Q_i \gg Q_e $, at zero magnetic field; $Q_i$ reduces with increasing magnetic field due to vortex induced losses and eventually goes to undercoupled regime, $Q_i < Q_e$; in undercoupled regime $\lvert S_{11}\rvert^2$ shows a dip at resonance frequency, while for $Q_i \gg Q_e $, change in $\lvert S_{11}\rvert^2$ is minimal. Although $Q_i$ can be tracked up to large magnetic field range using this choice of $Q_e$, it gives rise to high error-bars in fits for $Q_i$ when $Q_i \gg Q_e $ (near zero field). Another set of resonators with $Q_i \lesssim Q_e$ are studied to accurately determine $Q_i$ near zero field. Dependencies at lower fields are more accurately calculated using these undercoupled resonators having film thickness 80 nm and are shown in (e) and (f) for parallel and perpendicular magnetic fields respectively; here the error-bars have reduced to size comparable to the markers. Measurements were performed at $20\ \textrm{mK}$ temperature.}
\end{figure*}

Ratio of reflected to input signal for a one-port capacitively coupled resonator in reflection geometry, as seen in Fig.~\ref{fig:figure2} (a), is described by the reflection coefficient $S_{11}$. Within the linear response of the resonator, $S_{11}$ as a function of frequency, $f$ ($=\frac{\omega}{2\pi}$), for a half-wave single port resonator can be modeled by
\begin{equation} \label{eq1}
\begin{split}
S_{11}(\omega) & = 1-\frac{Q_{i}}{\frac{1}{2}(Q_{i}+Q_{e})+iQ_{i}Q_{e}\frac{\omega-\omega_0}{\omega_{0}}}
\end{split}
\end{equation}
where $Q_i$ is the internal quality factor, $Q_e$ is the external quality factor and $f_0\ (=\frac{\omega_0}{2\pi})$ is the resonance frequency. \cite{singh_optomechanical_2014} Fig.~\ref{fig:figure2} (b) shows a representative measurement of the real part of $S_{11}$ taken at 20 mK and zero magnetic field, along with the fitted curve using Eq.~\ref{eq1}. This allows us to extract the $Q_i$ for various studies done in this work. We find maximum $Q_i \approx 22000$ at 20 mK temperature and zero magnetic field. We note that $Q_e$ determines the coupling of the resonator to the external measurement circuitry. We have studied resonators which are designed to be overcoupled ($Q_i>Q_e$) and undercoupled ($Q_i<Q_e$) at base temperature and zero magnetic field, to capture variation of $Q_i$ in regimes with different internal loss rates. We have used the in-phase response (real part of $S_{11}$) to extract out the resonator parameters (details in the supplementary material).

As our primary motive is to develop these resonators for magnon coupling experiments, \cite{huebl_high_2013,zhang_strongly_2014,tabuchi_hybridizing_2014} it is imperative to characterize their response in magnetic field. Although the resonance frequency of a magnon mode depends on the geometry and the material properties, usually a magnetic field of $\sim 100\ \textrm{mT}$ could be sufficient to obtain magnon modes near 5 GHz. \cite{zhang_strongly_2014,macneill_gigahertz_2019,kapoor_observation_2021} Magnetic field dependence of the resonators in orientations parallel ($B_\textrm{para}$) and perpendicular ($B_\textrm{perp}$) to the SCPW plane has been shown in Fig.~\ref{fig:figure3}. Fig.~\ref{fig:figure3} (a) and (b) show the color-scale plots of $\lvert S_{11}\rvert ^2$ as a function of frequency and magnetic field for an NbN SCPW resonator, with film thickness 72 nm, on intrinsic silicon substrate, for field orientations parallel and perpendicular to the SCPW plane. These measurements have been done in zero-field-cooled condition to address magnetic field induced losses while continuously varying magnetic field, as is required in many applications. Note that here the cavity is designed to be overcoupled at zero magnetic field. In overcoupled regime, $\lvert S_{11}\rvert ^2$ shows small variations (however the resonance feature can be clearly seen in the phase response) and hence resonance dip is not visible at zero magnetic field. The dip starts appearing as vortex induced losses show up at higher fields.

Fig.~\ref{fig:figure3} (c) and (d) show the $Q_i$ values extracted from fit using Eq.~\ref{eq1} for resonators in magnetic field orientation parallel and perpendicular to the SCPW plane respectively. Resonators in parallel field show slower decrease in $Q_i$. This agrees with the fact that all the films have thicknesses below penetration depth. Whereas, for perpendicular field the rate of degradation of $Q_i$ is much faster due to the dominant role of vortex dynamics induced losses over quasiparticle losses. \cite{kwon_magnetic_2018}

Comparing the data for variation with magnetic field parallel to the SCPW plane, shown in Fig.~\ref{fig:figure3} (c) for 217~nm and 72~nm thick NbN films, we note that the degradation of $Q_i$ is slower for the thinner film as dissipation due to vortex dynamics plays a lesser role and only quasiparticle losses are important; while for the thicker film this is not the case and there is an early onset of vortex dynamics induced losses. NbN films with smaller thicknesses maintain higher $Q_i$ ($>10^3$) up to higher parallel magnetic field $B_\textrm{para} \approx1\ \textrm{T}$ due to lower flux creep. For orientation of magnetic field perpendicular to the plane of the film, $Q_i$ $>10^3$ is observed up to $B_\textrm{perp} \approx 100\ \textrm{mT}$ field.

In Fig.~\ref{fig:figure3} (c) and (d) the resonators are overcoupled ($Q_i \gg Q_e$) at zero field. The choice of $Q_e$ for this is good for tracking $Q_i$ up to larger field range, but leads to high error-bars in fits for $Q_i$ in regime of low internal loss, which is realized near zero magnetic field. So, for determining $Q_i$ more precisely in low magnetic field regime, we fabricated another set of resonators with lower coupling capacitor such that $Q_i \lesssim Q_e$ at zero field. Fig.~\ref{fig:figure3} (e) and (f) show the result of analysis for $Q_i$ with this $Q_e$ and we notice that these results are consistent and have smaller fitting errors (additional details in the supplementary material).

The near quadratic and linear dispersion trends in Fig.~\ref{fig:figure3} (a) and (b), in orientations parallel and perpendicular to the magnetic field respectively, is well described by Abrikosov-Gor'kov (AG) theory. \cite{tinkham_introduction_2004} For SCPW resonators, resonance frequency of the fundamental mode can be written as $f_0=\frac{\beta}{\sqrt{L_l}}$, where $\beta=\frac{1}{2l\sqrt{C_l}}$ with $l$, $C_l$ and $L_l$ being the length, capacitance per unit length and inductance per unit length of the resonator. The total inductance contains contribution from the geometric inductance ($L_g$, per unit length) and the kinetic inductance ($L_k$, per unit length). As the shift in resonance frequency comes from the kinetic inductance part, it is straight forward to obtain $\frac{\Delta L_k}{L_k}=-\frac{2\beta^2}{\beta^2-f_0^2L_g}\frac{\Delta f_0}{f_{0,0}}$ where $\Delta$ represents change in associated quantity from its $B=0\ \textrm{T}$ value (here $\Delta f_0$ is shift in resonance frequency from its value at base temperature and zero magnetic field,  $f_{0,0}$). Now we use the fact that for $T \ll T_C$, $L_k\propto\frac{1}{T_C}$ and $k_B\Delta T_C=-\frac{\pi \alpha}{4}$ where $\alpha$ is the half of depairing energy and for a thin film in perpendicular magnetic field case it can be written as $\alpha=DeB_\textrm{perp}$ where $D$ is the electronic diffusion constant and $e$ is magnitude of electron charge. \cite{tinkham_introduction_2004,samkharadze_high-kinetic-inductance_2016} Using this formalism we fit a straight line $\frac{\Delta f_0}{f_{0,0}}=-kB_\textrm{perp}$ to the dispersion (details in the supplementary material) and find $D$ from the relation $D=\frac{8k}{\pi e}\frac{\beta^2 k_B T_C}{\beta^2 - f_0^2 L_g}$. We find $D\approx5\times10^{-4}\ \textrm{m}^2\textrm{s}^{-1}$ which is close to values reported previously. \cite{mondal_enhancement_2013}

We observe a weak hysteresis in the measurement depending on the sweep direction of the magnetic field (details in the supplementary material). All the measurements presented in Fig.~\ref{fig:figure3} were recorded consistently with an upward sweep direction of the magnetic field. In parallel field orientation, a misalignment between the direction  of the magnetic field and the plane of the superconducting film, can also cause losses due to a non-zero out of plane component of the magnetic field. In our experiment, we estimated this misalignment to be less than $1\degree$, as the shift in resonant frequency remains flat for parallel magnetic field up to $\approx$ 0.8~T. We also characterized the resonators as a function of temperature and microwave power, and the details are provided in the supplementary material.

\begin{figure*}
\includegraphics{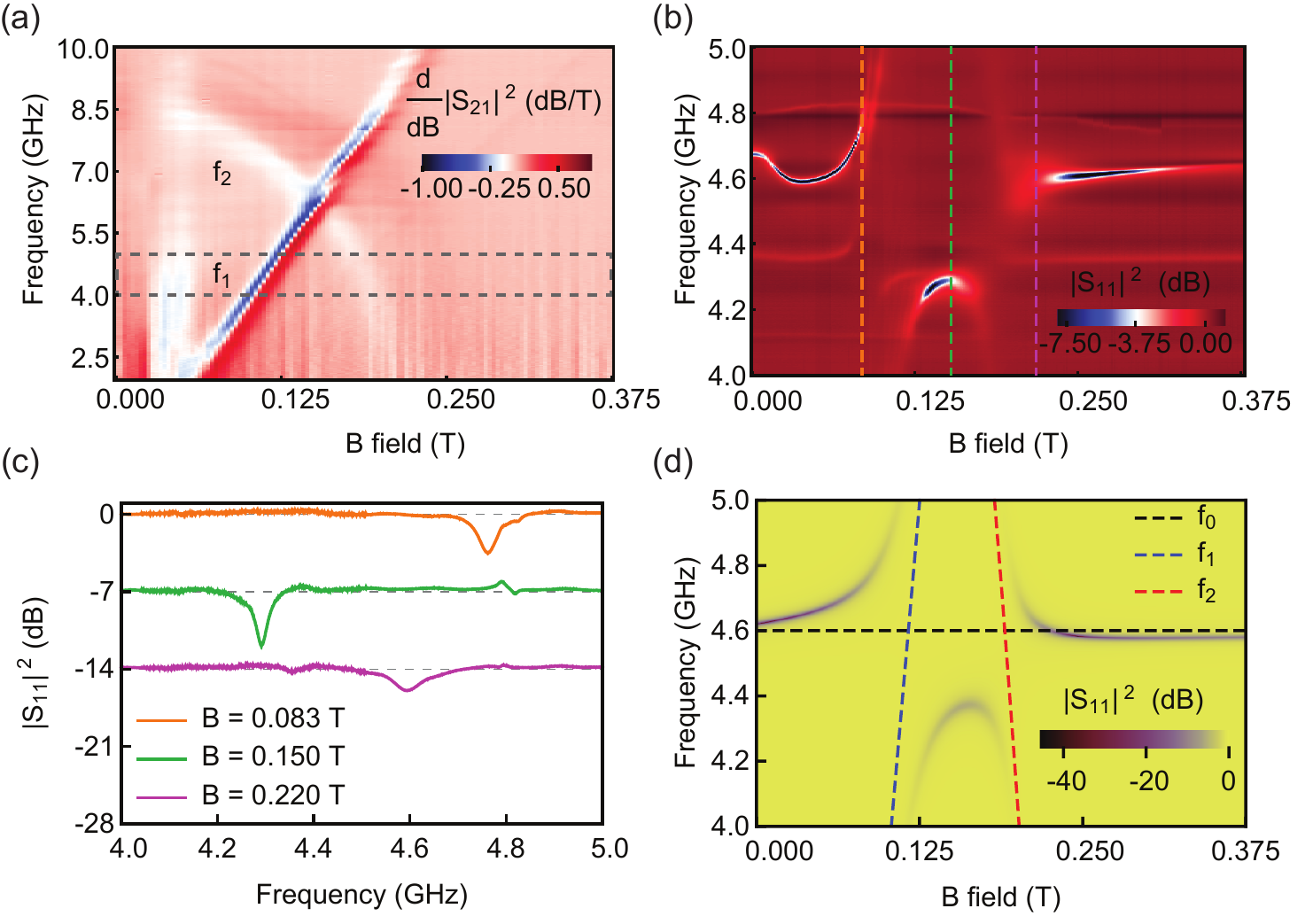}
\caption{\label{fig:figure4} Strong coupling between magnon modes in CrCl$_3$ and photon mode in NbN resonator. (a) Transmission spectra showing magnetic field dispersion of acoustic ($f_\textrm{1}$) and optical ($f_\textrm{2}$) AFMR modes in a CrCl$_3$ crystal flake placed on a copper coplanar waveguide transmission line. (b) Color-scale plot of the normalized $\lvert S_{11} \rvert^2$ with magnetic field and frequency for a CrCl$_3$ crystal flake placed at current antinode of an NbN SCPW resonator. The plot range is same as the gray dashed rectangle in (a) and shows the hybridization of the cavity mode with the acoustic and optical AFMR modes in CrCl$_3$ due to strong magnon-photon coupling. (c) Line-plots at three different fields (corresponding to dashed lines of same color in (b)). Offset of -7 dB has been added in consecutive plots for clarity. Line-shapes of upper hybrid modes, formed due to hybridization of cavity mode with acoustic and optical AFMR modes (orange and magenta, respectively), agree with the relatively broader line-width of optical mode compared to acoustic mode, as apparent from (a). (d) Plot of calculated $\lvert S_\textrm{11} \rvert^2$ using input-output theory with coupling strength of acoustic and optical AFMR modes with cavity mode of 0.57 GHz and 0.37 GHz respectively (dashed lines show the bare cavity ($f_\textrm{0}$), acoustic ($f_\textrm{1}$) and optical ($f_\textrm{2}$) AFMR modes).}
\end{figure*}

For application of the NbN SCPW resonators towards cavity magnonics devices, we investigated their coupling with magnons in CrCl$_3$ crystal, a van der Waals antiferromagnet. First, the microwave absorption spectroscopy of CrCl$_3$ was performed using a 20-30 µm thick CrCl$_3$ crystal flake placed on a coplanar waveguide type transmission line made out of copper on a printed circuit board (PCB) (measurement setup shown in supplementary material). The crystal was transferred to the PCB inside a glove-box and covered with Apeizon N grease to protect against ambient. This measurement was done in a cryostat under continuous flow of helium vapor, at a temperature of 1.5 K. On applying a magnetic field oriented parallel to the plane of the PCB and perpendicular to the direction of RF current in the transmission line, we observe two symmetry protected antiferromagnetic resonance modes with presence of a mode-crossing in the transmission spectra as shown in Fig.~\ref{fig:figure4}(a). The linearly dispersing mode ($f_1$) has been identified as the acoustic mode and the downward-going mode ($f_2$) has been identified as the optical mode. The two modes get excited when certain symmetry rules are satisfied. These modes and the symmetry arguments have been studied in detail in ref [\onlinecite{macneill_gigahertz_2019}] and [\onlinecite{kapoor_observation_2021}]. In our experiments, the DC magnetic field is always applied in the plane of the sample and perpendicular to the direction of the RF current. \cite{macneill_gigahertz_2019,kapoor_observation_2021} Using the derivative divide method\cite{maier-flaig_note_2018}, the data in Fig.~\ref{fig:figure4}(a) has been analyzed and the linewidth of the acoustic and optical modes have been determined as $\frac{\gamma_1}{2\pi}\approx$ 0.33 GHz and $\frac{\gamma_2}{2\pi}\approx$ 0.42 GHz respectively. The variation in these linewidths with magnetic field are found to be small.\cite{kapoor_observation_2021} We note that the optical mode has a broader linewidth compared to the acoustic mode.

After this, we placed a similar crystal flake at the current antinode of the NbN resonator on intrinsic Si chip inside glove-box and similarly covered with Apiezon N grease. This measurement was done in the dilution fridge and the grease provided additional benefit of good thermal anchoring of crystal to the 20 mK plate temperature. On applying magnetic field parallel to SCPW plane and perpendicular to the RF current direction at the current antinode, we observe formation of avoided crossing between the cavity mode and each of the two AFMR modes in the measured $\lvert S_\textrm{11} \rvert^2$, as shown in Fig.\ref{fig:figure4} (b). This is the signature of strong magnon-photon coupling between microwave photons in the NbN resonator and antiferromagnetic magnons in CrCl$_3$. Fig.~\ref{fig:figure4} (c) shows line-plots at different magnetic fields. Frequency dispersion of the modes in Fig.~\ref{fig:figure4} (b) and broader linewidth of the hybrid mode associated with the optical AFMR mode in Fig.~\ref{fig:figure4} (c) agree with the results from the transmission spectroscopy measurement using the copper transmission line.

We further generated a plot of $\lvert S_\textrm{11} \rvert^2$ from calculation based on input-output theory. \cite{schuster_high-cooperativity_2010,abe_electron_2011,clerk_introduction_2010} According to this theory, the reflection coefficient ($S_\textrm{11}$) for the system with two non-interacting magnon modes coupled with a cavity photon mode can be written as
\begin{equation} \label{eq2}
\begin{split}
S_{11}(\omega) & = 1-\frac{\kappa_{e}}{i(\omega-\omega_0)+\frac{\kappa_i+\kappa_e}{2}+\frac{\lvert g_1 \rvert^2}{i(\omega-\omega_1)+\frac{\gamma_1}{2}}+\frac{\lvert g_2 \rvert^2}{i(\omega-\omega_2)+\frac{\gamma_2}{2}}}
\end{split}
\end{equation}
where $\kappa_i=\frac{\omega_0}{Q_i}$ and $\kappa_e=\frac{\omega_0}{Q_e}$, $\frac{\omega_1}{2\pi}=f_1$ and $\frac{\omega_2}{2\pi}=f_2$ are the field dependent resonance frequencies of the acoustic and optical magnon modes respectively, $\frac{\gamma_1}{2\pi}$ and $\frac{\gamma_2}{2\pi}$ are their linewidths, and $\frac{g_1}{2\pi}$ and $\frac{g_2}{2\pi}$ are their coupling strength with the cavity mode respectively. For fitting the mode-coupling data shown in Fig.~\ref{fig:figure4} (b), we defined a three mode coupling Hamiltonian matrix. The diagonal elements of this symmetric matrix comprises of the bare cavity mode, the bare acoustic magnon mode and the bare optical magnon mode. We keep the off-diagonal terms representing the coupling between cavity and acoustic mode, and between cavity and optical mode as fitting parameters, and take the cross coupling term between the two magnon modes to be zero. The field dependence of the acoustic and the optical modes were obtained from polynomial fits to the bare magnon modes data shown in Fig.~\ref{fig:figure4} (a), and the cavity mode was assumed to be constant in the field range considered. We use the field dependent functional forms of the eigenmodes of this matrix to perform a non-linear model fitting to the experimentally obtained mode-coupling data shown in Fig.~\ref{fig:figure4} (d). This fitting gives $\frac{g_1}{2\pi}=0.57\ \textrm{GHz}$ and $\frac{g_2}{2\pi}=0.37\ \textrm{GHz}$. Using these $g$ values, we obtain the plot as shown in Fig.~\ref{fig:figure4} (d) with the expression from input-output theory Eq.~\ref{eq2}. The calculated cooperativity of coupling between acoustic magnon and photon, and between optical magnon and photon, using these values are $C_1=\frac{g_1^2}{\kappa \gamma_1}=47.10$ and $C_2=\frac{g_2^2}{\kappa \gamma_2}=15.59$ respectively, where $\kappa=\kappa_i+\kappa_e$. The cooperativities larger than the expected modes occupancy clearly suggest a quantum-coherent coupling between the photons and the magnons.  

Since we use a CrCl$_3$ crystal covering almost the entire SCPW resonator, the effective volume for magnon-photon coupling is determined mainly by the magnetic mode volume of the resonator. From the lattice structure parameters of CrCl$_3$ and using the magnetic component of the vacuum RF field of the resonator\cite{huebl_high_2013,lambert_identification_2015}, we calculate the coupling strength per spin to be $\frac{g_0}{2\pi}\approx 7.93$ Hz and a net coupling strength of $\approx 1.09$ GHz for the estimated effective volume (more details regarding fitting and $g$ value estimation is provided in the supplementary material). The deviation in the estimated and experimentally measured coupling rates could be attributed to reduction in the effective volume as predominantly the spins near the gap of resonator contribute to the measured signal. Also, coupling strength for the acoustic and the optical modes separately depend on both geometry and symmetry, as we observe them to be.  A more accurate microscopic calculation and further experiments are needed to fully understand some aspects of the mode coupling data. This can be a focus of our future work. We also observe a faint dispersing mode below the main cavity mode in Fig.~\ref{fig:figure4} (b) which also couples with the AFMR modes. This is possibly a parasitic mode of the resonator whose origin needs to be verified. Furthermore, a slight downward shift in cavity frequency with increasing magnetic field, seen between 10~mT and 30~mT, could possibly be attributed to the change in magnetic state of CrCl$_3$ as it undergoes spin-flop transition. \cite{narath_spin-wave_1965,mcguire_magnetic_2017,kuhlow_magnetic_1982} This measurement shows that our NbN resonators can be an ideal platform for studying collective spin oscillations and their coupling with electromagnetic modes of microwave frequencies in different systems.

In this work, we have fabricated NbN SCPW resonators using a simple fabrication process and probed their baseline properties. We see $Q_i>10^3$ persisting up to perpendicular magnetic field of 100 mT which is two times higher than the previously reported results. \cite{kroll_magnetic-field-resilient_2019} Using substrate surface treatments and vortex trapping schemes $Q_i$ of these resonators can be made even higher with better performance possibly up to even higher magnetic fields. Our fabrication protocols have the potential to incorporate exfoliable crystals in the microwave circuits. Furthermore, we have demonstrated the effectiveness of these resonators in coupling with spin ensembles by coupling them to magnons in a van der Waals antiferromagnetic system to study collective spin oscillations and making hybrid quantum devices.

\paragraph*{\\}
See the supplementary material for details on measurement circuit, fitting procedure along with comparison between overcoupled and undercoupled resonators, characterization data for resonators on intrinsic silicon and sapphire substrate with temperature and microwave power, and details about magnon-photon coupling experiment.

\paragraph*{\\}
We thank Rajamani Vijayaraghavan, Meghan Patankar, Ipsita Das, Sudhir Sahu, Suman Kundu, Sumeru Hazra, Ameya Riswadkar, Anirban Bhattacharjee and Srijita Das for helpful discussion and experimental assistance. We also thank Bhagyashree Chalke and Rudheer Bapat for doing SEM imaging.  We acknowledge the Swarnajayanti Fellowship of the Department of Science and Technology (for M.M.D.), DST Nanomission grant SR/NM/NS-45/2016, SERB SUPRA SPR/2019/001247, ONRG grant N62909-18-1-2058, and the Department of Atomic Energy of the Government of India grant 12-R\&D-TFR-5.10-0100 for support.

\onecolumngrid

\section*{Supplementary Material}

\renewcommand{\theHsection}{Ssection.\thesection}
\renewcommand{\thesection}{S\Roman{section}}
\setcounter{section}{0}

\renewcommand{\theHfigure}{Sfigure.\thefigure}
\renewcommand{\thefigure}{S\arabic{figure}}
\setcounter{figure}{0}

\renewcommand{\theHequation}{Sequation.\theequation}
\renewcommand{\theequation}{S\arabic{equation}}
\setcounter{equation}{0}

\renewcommand{\theHtable}{Stable.\thetable}
\renewcommand{\thetable}{S\Roman{table}}
\setcounter{table}{0}

\section{Schematic for measurement circuit}
\begin{figure*}[h]
\includegraphics{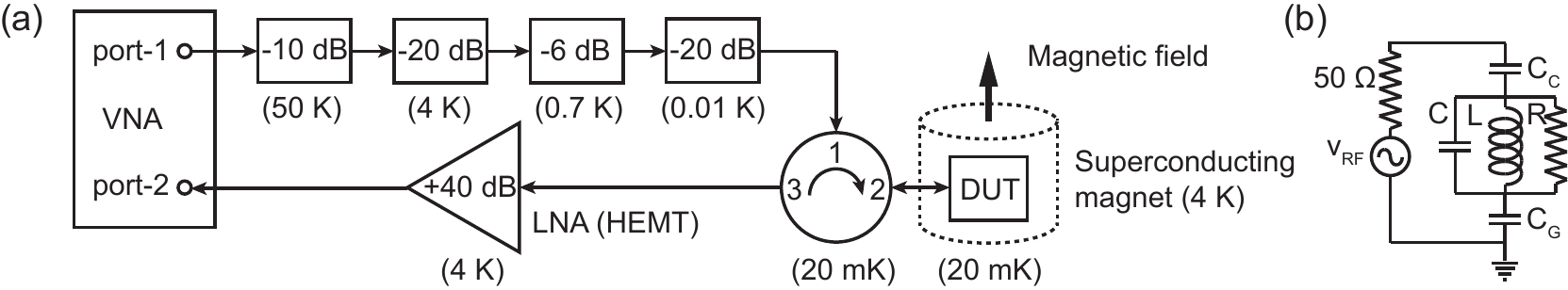}
\caption{\label{fig:fig1} (a) Circuit schematic for reflection measurement of NbN SCPW resonators in an Oxford Triton dilution fridge. RF signal from one port of a VNA is carried through a coax line with attenuation at different temperature plates inside the fridge and to the device under test (DUT) through a circulator. Reflected signal from the device is amplified by a Low Noise Factory HEMT amplifier at 4 K plate before sending it back to the other port of the VNA. (b) A lumped element equivalent of the device. Here the parallel combination of $L$, $C$ and $R$ represent the resonator, $C_C$ represents the coupling capacitor, $C_G$ is a capacitor representing the open-to-ground configuration of the one-port resonator and $v_\textrm{RF}$ represents the VNA, which is used for actuation and measurement of the resonator.}
\end{figure*}
Fig.~\ref{fig:fig1} (a) shows schematic of the circuit for reflection measurement of the NbN SCPW resonators. Measurements are done in an Oxford Triton dilution fridge containing a superconducting magnet. Input line of the RF signal includes several attenuators at different stages of the fridge as shown, for proper thermalization of the microwave photons reaching the sample. Input and output paths are separated using a circulator from Quinstar (OXE89 CTH0408KC) before the device. Amplification is done by a Low Noise Factory amplifier (LNF-LNC0.3\_14A) attached to the 4 K plate of the fridge.  Attained base temperature at sample is 20 mK. The temperature of the sample varies a bit 10-20 mK, so we use the upper range of the temperature. The superconducting coil around the sample is used to apply magnetic field. The measurements are done using an Anritsu (MS46122B) vector network analyzer (VNA). Fig.~\ref{fig:fig1} (b) shows a lumped element equivalent to the device. Note that, because of the signal propagating through the cables, an electrical length dependent phase factor gets multiplied to $S_\textrm{11}(f)$ from the device, giving a modified reflectivity $S_\textrm{11,mod}(f)=S_\textrm{11}(f)\exp{(i\ 2\pi f \frac{L}{v_p})}$ where $L$ is the effective length traversed by the signal and $v_p$ is the effective phase velocity of the signal. This electrical length dependent phase part, being a monotonic function of frequency, adds a slope to the background of the phase of the measured signal. Hence, this shows up in the plots of real and imaginary parts of $S_\textrm{11}$ as well. By correcting for this overall slope in the background, the pure device response is extracted. We use the functionality available in the VNA to correct for the electrical length while recording data.

\section{Comparison between overcoupled and undercoupled resonators}
\begin{figure*}[h]
\includegraphics{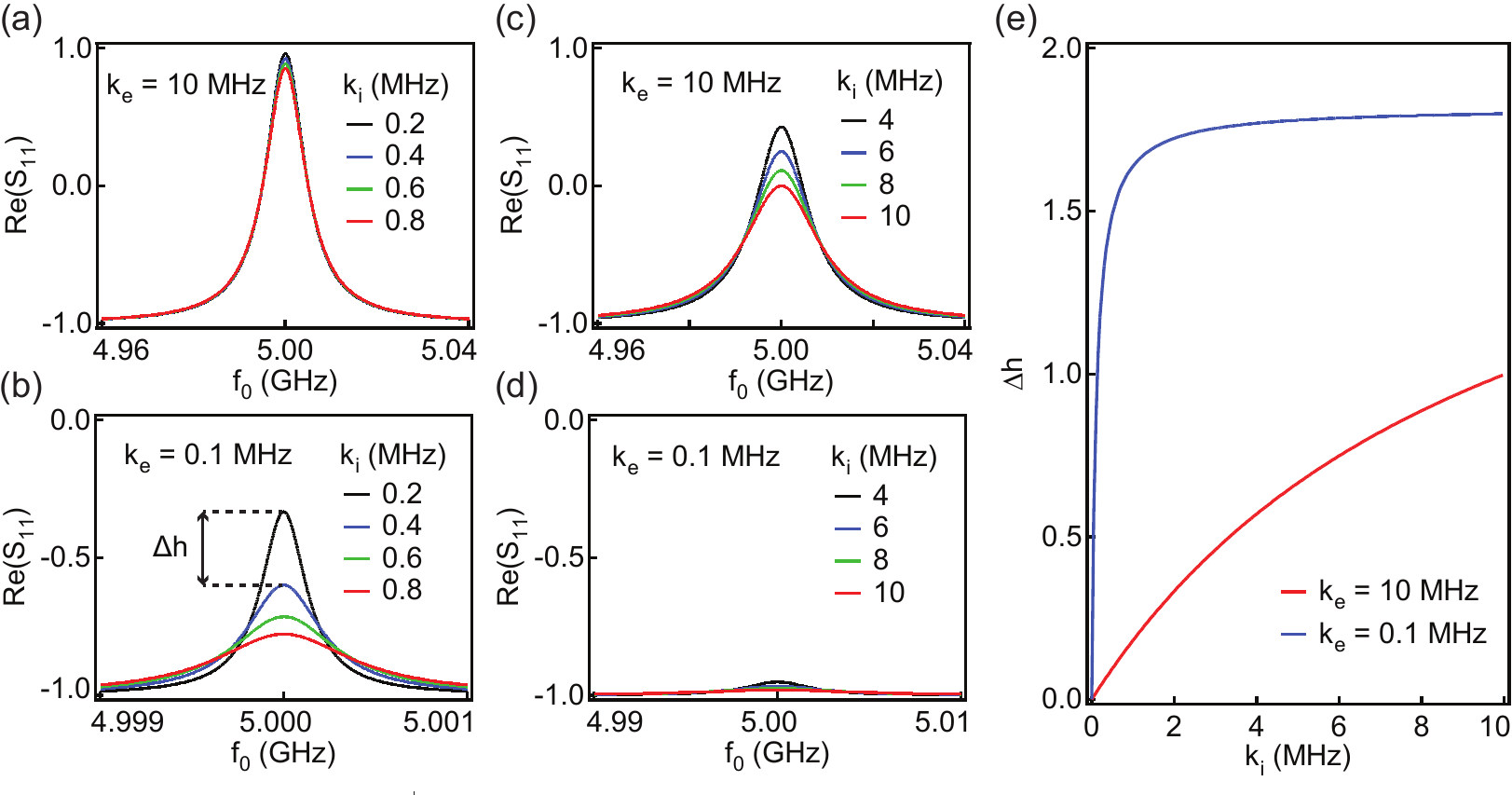}
\caption{\label{fig:fig2} Numeric calculations showing shift in peak height $\Delta h$ of real part of $S_{11}$. (a) and (b) show $\Delta h$ in high $Q_i$ (low $k _i$) regime for resonator designs with low ($k_e=10\ \textrm{MHz}$) and high ($k_e=0.1\ \textrm{MHz}$) $Q_e$ values respectively (i.e. overcoupled and undercoupled respectively). (c) and (d) show $\Delta h$ in low $Q_i$ (low $k_i$) regime for resonators with these two designs respectively. (e) shows a comparison between these two designs of $k_e$ at different $k_i$. From these plots we observe that an overcoupled design is preferable for good fit in low $Q_i$ regime and an undercoupled design is preferable for good fit in high $Q_i$ regime.}
\end{figure*}
Proper designing of coupling capacitor ($C_C$) is essential in extracting internal $Q$ of the resonators accurately in regimes with different internal loss rates. Reflection coefficient ($S_{11}$) of a one-port resonator as a function of frequency is given by
\begin{equation} \label{eq1}
\begin{split}
S_{11}(f) & = 1-\frac{k_{e}}{\frac{k_{i}+k_{e}}{2}+i {(f-f_0)}}
\end{split}
\end{equation}
where $k_i=\kappa_i/2\pi$ and $k_e=\kappa_e/2\pi$ are internal and external loss rates respectively and $f_0=\omega_0/2\pi$ is the resonance frequency of the resonator. \cite{singh_optomechanical_2014} They are related to internal $Q$ ($Q_i$) and external $Q$ ($Q_e$) by $Q_i=\frac{\omega_0}{\kappa_i}=\frac{f_0}{k_i}$ and $Q_e=\frac{\omega_0}{\kappa_e}=\frac{f_0}{k_e}$. Note that an overall constant phase factor to $S_{11}$ with 0 or $\pi$ phase can give rise to a dip or peak respectively in the real part of $S_{11}$, which are equivalent.
For understanding effect of chosen $Q_e$ in accuracy of extracted $Q_i$ using Eq.~\eqref{eq1}, we do numeric calculations in Mathematica, as shown in Fig.~\ref{fig:fig2}. Greater shift in peak height ($\Delta h$) of real part of $S_{11}$ for same $Q_i$ variation provides higher accuracy in estimation of $Q_i$. Fig.~\ref{fig:fig2} (a) and (b) show variation in $\Delta h$ in high $Q_i$ (i.e. low $k_i$) regime for high (10 MHz) and low (0.1 MHz) values of chosen $k_e$ respectively. For the higher values of $Q_i$ considered in Fig.~\ref{fig:fig2}, these correspond to overcoupled ($Q_i>Q_e$) and undercoupled ($Q_i<Q_e$) designs respectively. High $Q_i$ regime is realized near base temperature of 20 mK and zero magnetic field. Fig.~\ref{fig:fig2} (c) and (d) show variation in $\Delta h$ in low $Q_i$ (i.e. high $k_i$) regime for these two choices of $k_e$. Low $Q_i$ regime is realized as losses are introduced in the system due to increase in temperature and/or magnetic field. Fig.~\ref{fig:fig2} (e) shows a comparison between $k_e=10\ \textrm{MHz}$ and $k_e=0.1\ \textrm{MHz}$ in terms of variation of $\Delta h$ with $k_i$. From these panels we note that in the high $Q_i$ regime undercoupled design (low $k_e$ with $k_e<k_i$) is preferable for good fit, whereas in the low $Q_i$ regime overcoupled design (high $k_e$ with $k_e>k_i$) is preferable for good fit. We use resonators with $Q_e \approx 400$ and $Q_e \approx 22000$ for characterization of the resonators. We find that although lower $Q_e$ is useful for extracting $Q_i$ up to higher magnetic fields it gives high errorbars in the lower magnetic fields. Whereas, using higher $Q_e$ enables us to extract $Q_i$ with higher accuracy in low magnetic field regime, as implied by much lower error bars as shown in main text (Fig.4). This agrees with our comparison in Fig.~\ref{fig:fig2}.

\section{Simultaneous fitting to real and imaginary parts of $S_{11}$}
\begin{figure*}[h]
\includegraphics{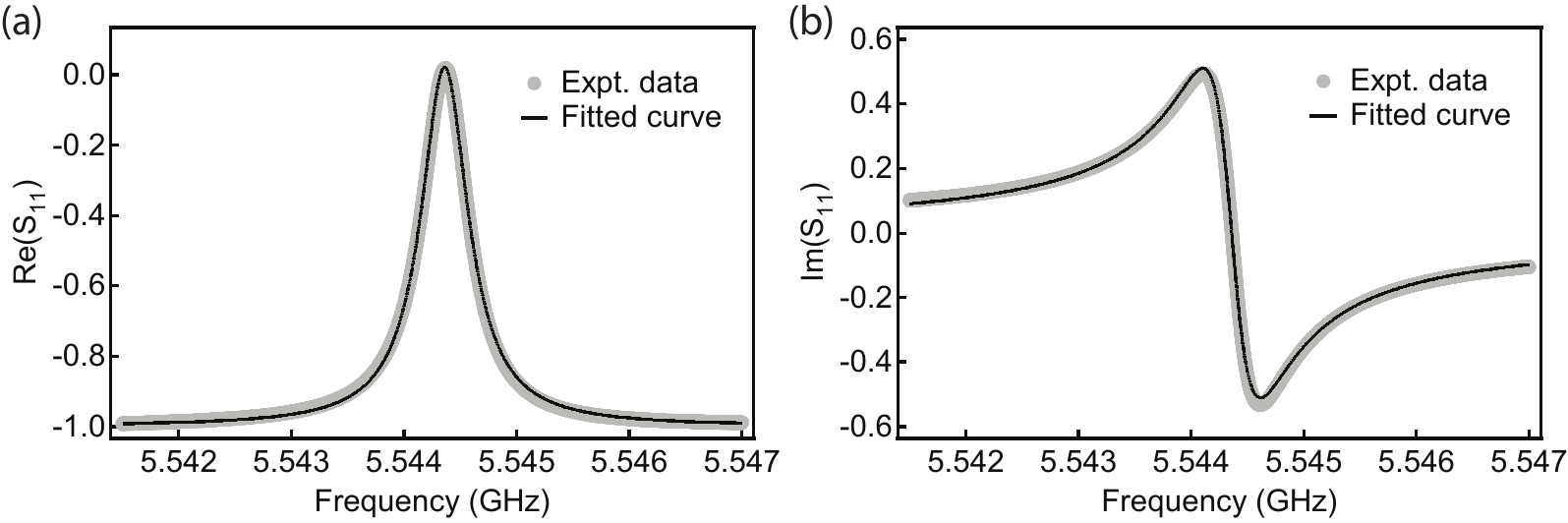}
\caption{\label{fig:fig3} (a) and (b) Simultaneous fit to real and imaginary parts of $S_{11}$ respectively, of normalized experimental data for an NbN SCPW resonator at zero magnetic field and 20 mK base temperature using Eq.~\ref{eq1}.}
\end{figure*}
We have extracted system parameters by fitting to the real part of $S_{11}$ data using Eq.~\ref{eq1}. Simultaneous fitting to both real and imaginary part of $S_{11}$ is also possible for slight improvement. Fig.~\ref{fig:fig3} (a) and (b) show the simultaneous fits to real and imaginary parts of normalized $S_{11}$ obtained from measurement of a resonator with $Q_e \approx Q_i \approx 22000$; this is close to the value obtained by fitting only to the real part of $S_{11}$ as mentioned in main text. Further improvement can be done by implementing ways of continuous background calibration during magnetic field sweep to eliminate the background, which shows slight variation with increasing magnetic field, from the measured signal.

\section{Comparison with previous reports}
\begin{table}[h]
\caption{Comparison of $Q_i$ variation with magnetic field with previous reports}
\label{table:1}
%\begin{adjustbox}{width=\columnwidth,center}
\begin{tabular}{|l|l|l|l|l|}
\hline \hline
Material & \makecell[l]{Resonator\\geometry} & \makecell[l]{Flux trapping\\scheme} & Max $B_\textrm{para}$ (for $Q_i>10^3$) & Max $B_\textrm{perp}$ (for $Q_i>10^3$) \\ \hline
Nb [\onlinecite{kwon_magnetic_2018}] & SCPW & No & 2.7 T (loaded $Q>10^3$) & 24 mT (loaded $Q>2.5 \times 10^4$) \\ \hline
NbTiN [\onlinecite{kroll_magnetic-field-resilient_2019}] & SCPW & Yes & 6 T ($Q_i>10^5$) & 45 mT (20 mT for $Q_i>10^5$) \\ \hline
NbTiN nanowire [\onlinecite{samkharadze_high-kinetic-inductance_2016}] & Nanowire & No & 6 T ($Q_i>2 \times 10^5$) & 400 mT ($Q_i>10^4$) \\ \hline
NbN (our work) & SCPW & No & 1 T & 100 mT \\ \hline \hline
\end{tabular}
%\end{adjustbox}
\end{table}
Table \ref{table:1} shows a comparison of highest applied magnetic field, in orientations parallel and perpendicular to the SCPW plane, up to which $Q_i>10^3$ is retained, with previous reports for other type-II superconductors. We observe that even in absence of any substrate surface treatment and flux trapping schemes, our NbN resonators retain $Q_i>10^3$ up to parallel fields comparable to and perpendicular field twice the maximum value reported previously for an SCPW resonator. Further implementation of substrate surface treatments and flux trapping schemes are expected to increase this maximum field of high $Q_i$ retention even more.

\begin{figure*}[h]
\includegraphics{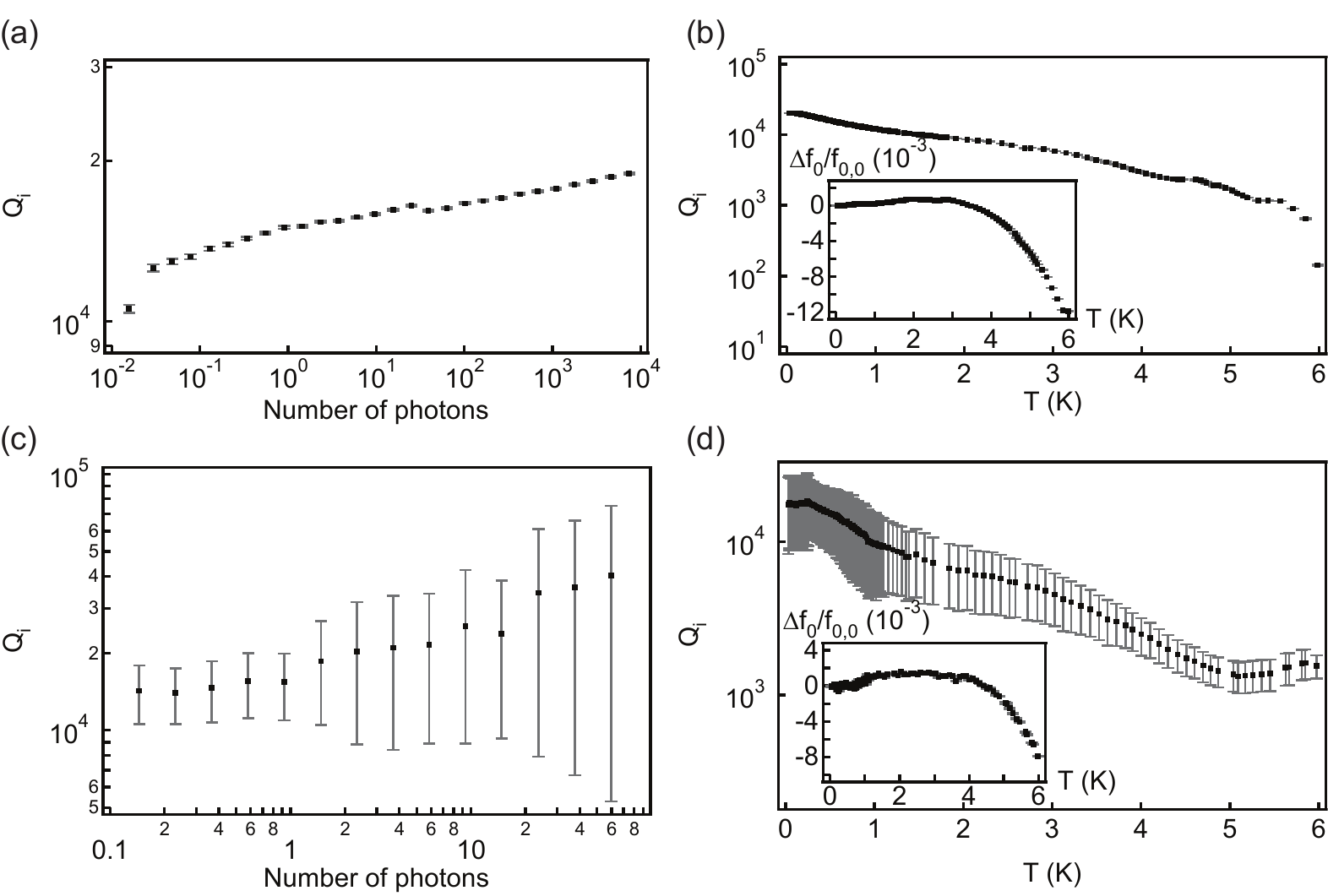}
\caption{\label{fig:fig4} Characterization of NbN SCPW resonators as a function of microwave power and temperature. (a) and (b) Variation of $Q_i$ with increasing number of microwave photons and temperature, respectively, for intrinsic silicon substrate [Inset of (b): Variation of $\Delta f_0/f_{0,0}$ with temperature for the same]. Note that the error-bars are of size comparable to the markers. (c) and (d) Same for resonators on a sapphire substrate.}
\end{figure*}

\section{Characterization of resonators with microwave power and temperature}

For characterizing the fabricated resonators with respect to experimental parameters, we study the internal loss of the resonators by varying experimental conditions such as drive power and temperature. Dependence of resonance characteristics on $P_\textrm{RF}$ is performed to investigate the performance of these resonators at and below single photon power level. The microwave power dependence of $Q_i$ is shown in Fig.~\ref{fig:fig4} (a) for intrinsic silicon substrate and in Fig.~\ref{fig:fig4} (c) for sapphire substrate. From Fig.~\ref{fig:fig4} (a) and (c) we observe a small increasing trend in $Q_i$ as the mean number of microwave photons in the cavity is increased. Such a behavior suggests the presence of two-level systems (TLS). \cite{gao_experimental_2008} At higher power, TLS get saturated and result in higher $Q_i$. For all the magnetic field dependent measurements discussed in the manuscript, we use a measurement power equivalent to approximately $10^3$ photons in the resonator. Temperature dependence of the $Q_i$ has been shown in Fig.~\ref{fig:fig4} (b) for intrinsic silicon substrate and in Fig.~\ref{fig:fig4} (d) for sapphire substrate (the insets show temperature dependence of $\frac{\Delta f_0}{f_{0,0}}$). As a general trend, above $T\approx4\ \textrm{K}$ $f_0$ of the resonators show a downward shift. This reduction in $f_0$ can be attributed to the increasing kinetic inductance due to reduction in the number density of available Cooper pairs. Furthermore, for the same reason, increase in number density of quasiparticles causes internal losses to increase, thereby lowering $Q_i$. A downward shift in the relative frequency shift below $T\approx1.5\ \textrm{K}$ confirms the presence of TLS. \cite{gao_experimental_2008}

\section{Characterization of resonators on sapphire substrate with magnetic field}
\begin{figure*}[h]
\includegraphics{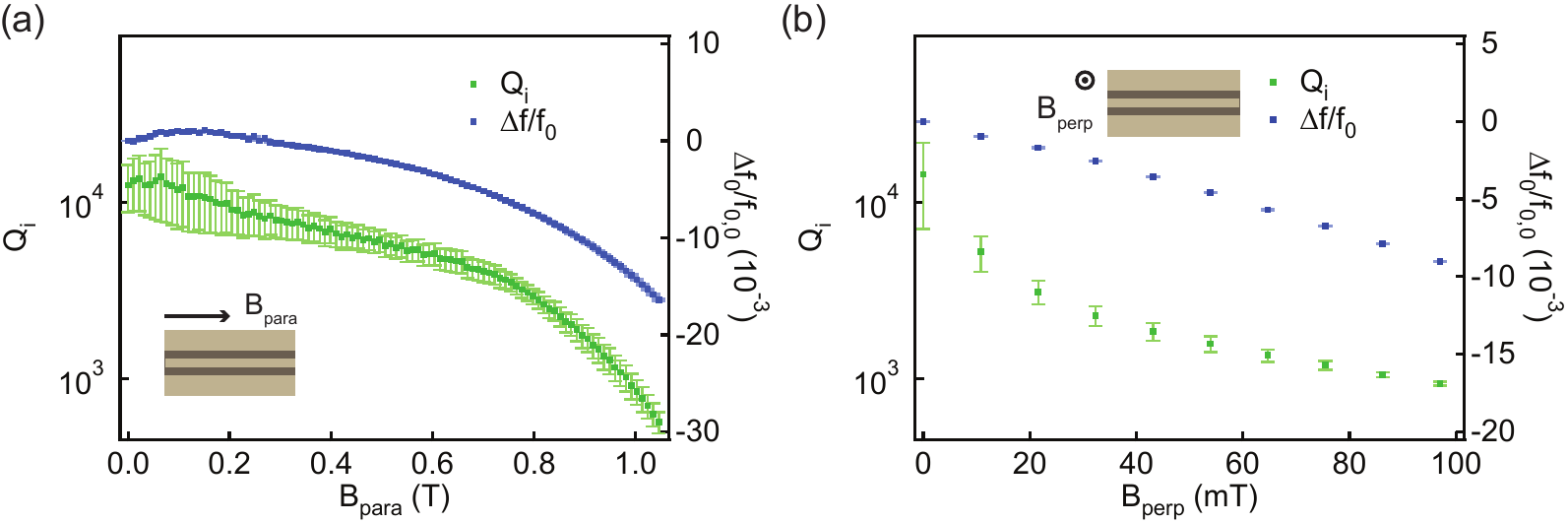}
\caption{\label{fig:fig5} Characterization of an NbN SCPW resonator on sapphire substrate (NbN film thickness 94 nm) as a function of magnetic field. (a) and (b) Variation of $Q_i$ and $\Delta f_0/f_{0,0}$ for parallel and perpendicular field orientations respectively.}
\end{figure*}
Fig.~\ref{fig:fig5} (a) and (b) show variation of $Q_i$ and $\Delta f_0/f_{0,0}$ with magnetic field parallel ($B_\textrm{para}$) and perpendicular ($B_\textrm{perp}$) to the SCPW plane respectively. This dependence is similar to that of resonators on intrinsic silicon substrate as shown in main text.

\section{Variation in $\Delta f_0/f_{0,0}$ with magnetic field for resonators on intrinsic silicon substrate}
\begin{figure*}[h]
\includegraphics{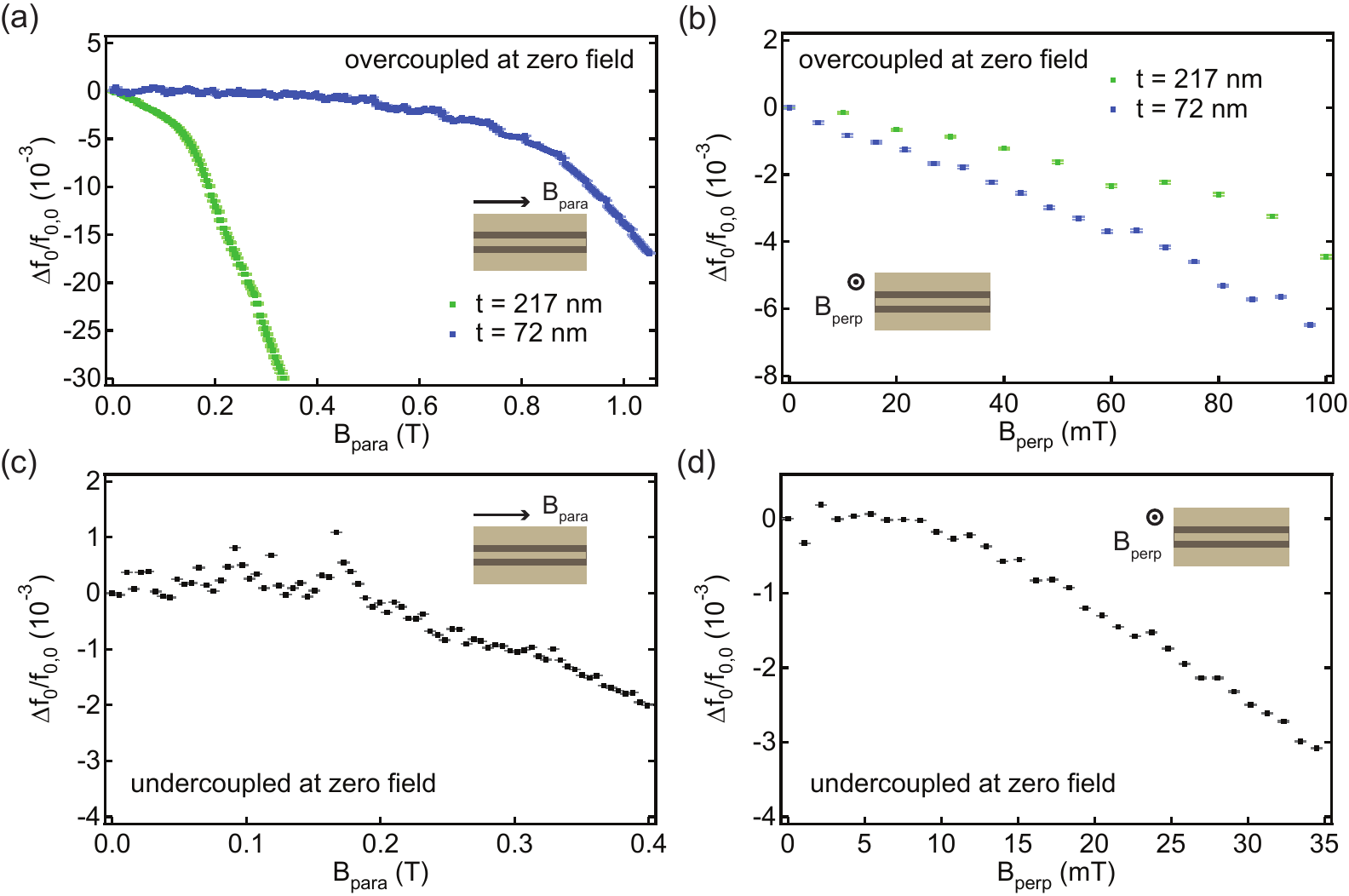}
\caption{\label{fig:fig6} Relative shift in resonance frequency $\Delta f_0/f_{0,0}$ of NbN resonators on intrinsic silicon substrate. (a) and (b) Variation in $\Delta f_0/f_{0,0}$ for the overcoupled resonators with NbN film thicknesses 217 nm and 72 nm in parallel and perpendicular magnetic field respectively. (c) and (d) Variation in $\Delta f_0/f_{0,0}$ for the undercoupled resonators with NbN film thickness 80 nm in parallel and perpendicular magnetic field respectively.}
\end{figure*}
Fig.~\ref{fig:fig6} (a) and (b) show the variation of $\Delta f_0/f_{0,0}$ for resonators on intrinsic silicon substrate with film thicknesses 217 nm and 72 nm with magnetic field parallel and perpendicular to the SCPW plane respectively. These show the nearly quadratic and linear dispersion of resonance frequency in parallel and perpendicular magnetic field respectively. The slope of the linear dispersion has been used to calculate electronic diffusion constant $D$. Fig.~\ref{fig:fig6} (c) and (d) show the same for resonators with undercoupled design.

\section{Hysteresis with magnetic field}
\begin{figure*}[h!]
\includegraphics{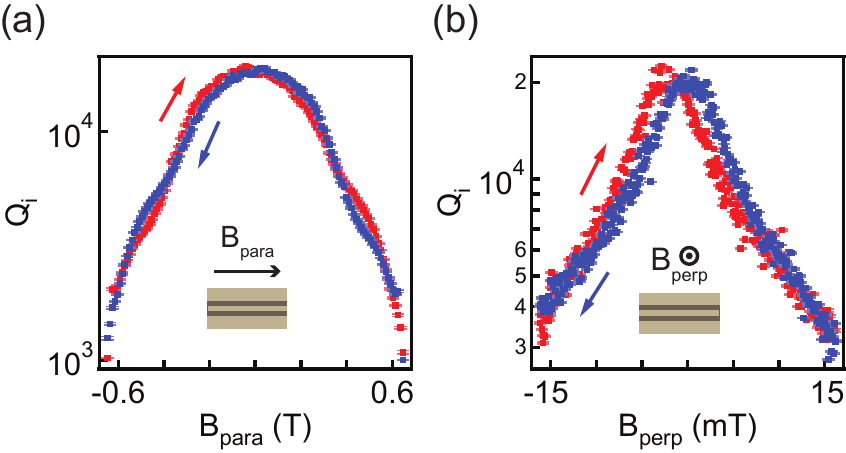}
\caption{\label{fig:fig7} Hysteresis in NbN SCPWs with magnetic field. (a) and (b) Hysteresis in an NbN SCPW resonator on intrinsic silicon substrate with respect to direction of magnetic field sweep, with field parallel and perpendicular to the SCPW respectively.}
\end{figure*}
We observe some amount of hysteresis in the NbN SCPW resonators depending upon  the direction of magnetic field sweep. Fig.~\ref{fig:fig7} (a) and (b) show hysteresis in $Q_i$ of a resonator on intrinsic silicon substrate. We observe that for the perpendicular field, $Q_i$ is higher in down-sweep of field compared to the up-sweep; whereas, for the parallel field, we observe regimes with higher as well as lower $Q_i$ in down-sweep compared to up-sweep. The perpendicular field case is similar to previous report and is similar to prediction from Norris-Brandt-Indenbom (NBI) model with inhomogeneous current density. \cite{bothner_magnetic_2012} But the presence of two regimes in case of parallel field sweep suggests the presence of grain boundaries \cite{ji_magnetic-field-dependent_1993,raychaudhuri_radio-frequency_1996} in the NbN film, which is also apparent from the SEM image of an NbN film as shown in the inset of Fig.1 in the main text.

\section{Linear fit for finding D}
\begin{figure*}[h]
\includegraphics{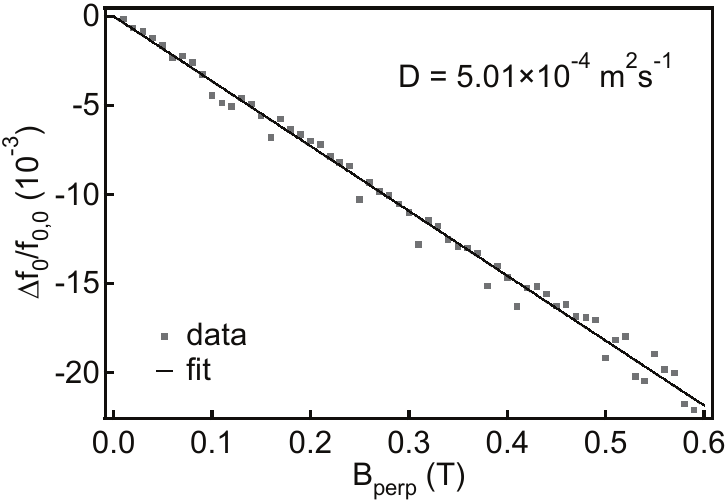}
\caption{\label{fig:fig8} Linear fit to the $\frac{\Delta f_0}{f_{0,0}}$ vs $B_\textrm{perp}$ data for determination of electronic diffusion constant ($D$) for an NbN SCPW resonator on intrinsic silicon substrate.}
\end{figure*}
The electronic diffusion constant ($D$) is given by $D=\frac{8k}{\pi e}\frac{\beta^2 k_B T_C}{\beta^2 - f_0^2 L_g}$ as described in main text, where $k$ is the negative of slope of $\frac{\Delta f_0}{f_{0,0}}$ vs $B_\textrm{perp}$ plot. Fig.~\ref{fig:fig8} shows a linear fit to the $\frac{\Delta f_0}{f_{0,0}}$ vs $B_\textrm{perp}$ data for an NbN SCPW resonator on intrinsic silicon substrate. The estimated electronic diffusion constant is $D=5.01\times10^{-4}\ \textrm{m}^2\textrm{s}^{-1}$.

\section{Transmission spectroscopy of chromium trichloride}

\begin{figure*}[h]
\includegraphics{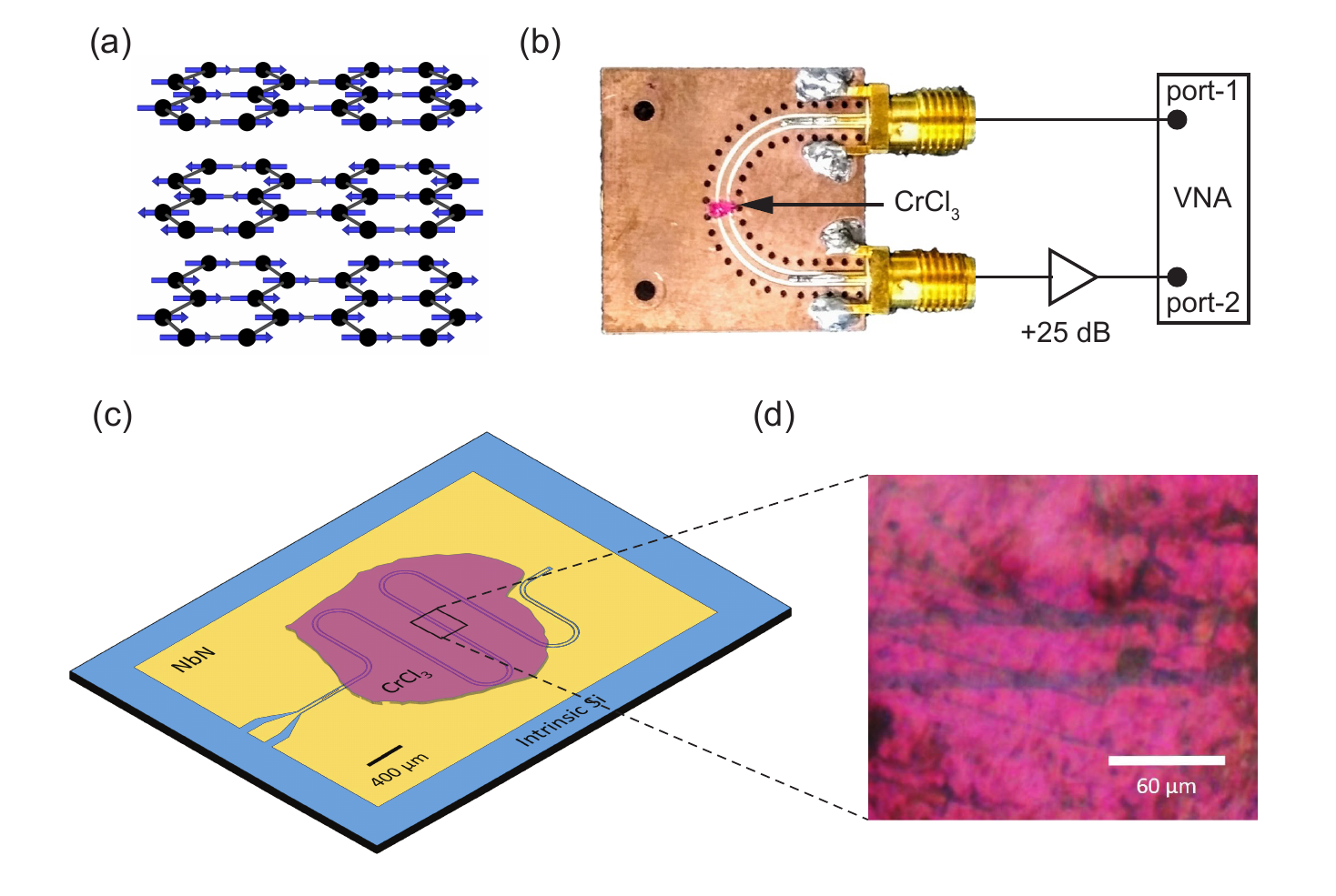}
\caption{\label{fig:fig9} (a) Crystal structure of CrCl$_3$ shows van der Waals stacking of antiferromagnetically oriented consecutive atomic layers (black spheres denote the Cr$^{3+}$ ions and blue arrows depict associated moments). (b) Measurement schematic for the transmission spectroscopy of a CrCl$_3$ crystal flake placed on a transmission line made of copper on a PCB. (c) Schematic showing a CrCl$_3$ crystal flake placed on the central part of an NbN SCPW resonator for the magnon-photon coupling measurement. (d) Optical microscopy image of the crystal in the part shown by a rectangle in (c).}
\end{figure*}

Chromium trichloride (CrCl$_3$) is a van der Waals antiferromagnet with weak interlayer exchange. Fig.~\ref{fig:fig9} (a) shows the crystal structure of CrCl$_3$. The localized magnetic moments in CrCl$_3$ are provided by the Cr$^{3+}$ ions shown as black spheres. In any atomic layer of CrCl$_3$, the magnetic moments are in plane, shown by blue arrows, and are coupled by ferromagnetic exchange. Magnetic moments in consecutive planes are antiferromagnetically coupled. Due to weak interlayer exchange of 1.6 µeV, antiferromagnetic resonance (AFMR) is observed at GHz frequencies in CrCl$_3$ whereas, most antiferromagnets have resonance modes in THz frequency range. This allows realization of coupling of the AFMR modes in CrCl$_3$ with cavity modes in coplanar waveguide resonators operating in the low GHz frequencies. 

To study the transmission spectra of CrCl$_3$, the crystal flake was placed on a copper CPW transmission line PCB and secured with Apezion grease to protect it from ambient. The measurement was done in a He flow cryostat. As shown in Fig.~\ref{fig:fig9} (b), using a vector network analyzer, microwave signal was sent from port-1 to the transmission line and the transmitted signal was amplified before being received at the port-2 of the VNA. DC magnetic field was applied parallel to the plane of the PCB and perpendicular to the segment of the transmission line containing the CrCl$_3$ crystal (i.e. perpendicular to the RF current at the position of the CrCl$_3$). The transmission spectra was obtained by measuring $S_\textrm{21}(f)$ at each value of DC magnetic field as it was swept up starting from zero field. The derivative of the transmission spectra with respect to the DC magnetic field was then plotted against the magnetic field and frequency, as shown in Fig. 4(a) of the manuscript. Using the derivative divide method,\cite{maier-flaig_note_2018} the background signal can be removed from the transmission spectra without calibration of the microwave circuits. This data can then be fit to an anti-Lorentzian to determine the linewidths of the modes. A detailed study of the different modes seen in CrCl$_3$ is presented in ref [\onlinecite{macneill_gigahertz_2019}] and [\onlinecite{kapoor_observation_2021}] . 

Fig.~\ref{fig:fig9} (c) shows the schematic of CrCl$_3$ crystal flake placed on an NbN SCPW resonator and Fig.~\ref{fig:fig9} (d) shows the optical microscopy image of the grease covered CrCl$_3$ crystal on the NbN resonator at the current antinode of the resonator (the part shown by a rectangle in Fig.~\ref{fig:fig9} (c)). As the crystal covers the entire NbN resonator, the effective volume of interaction between the microwave photons in the NbN resonator and the magnons in CrCl$_3$ is determined mainly by the mode volume of the magnetic component of the RF field of the resonator.

\section{Fit to magnon dispersion}
\begin{figure*}[h]
\includegraphics{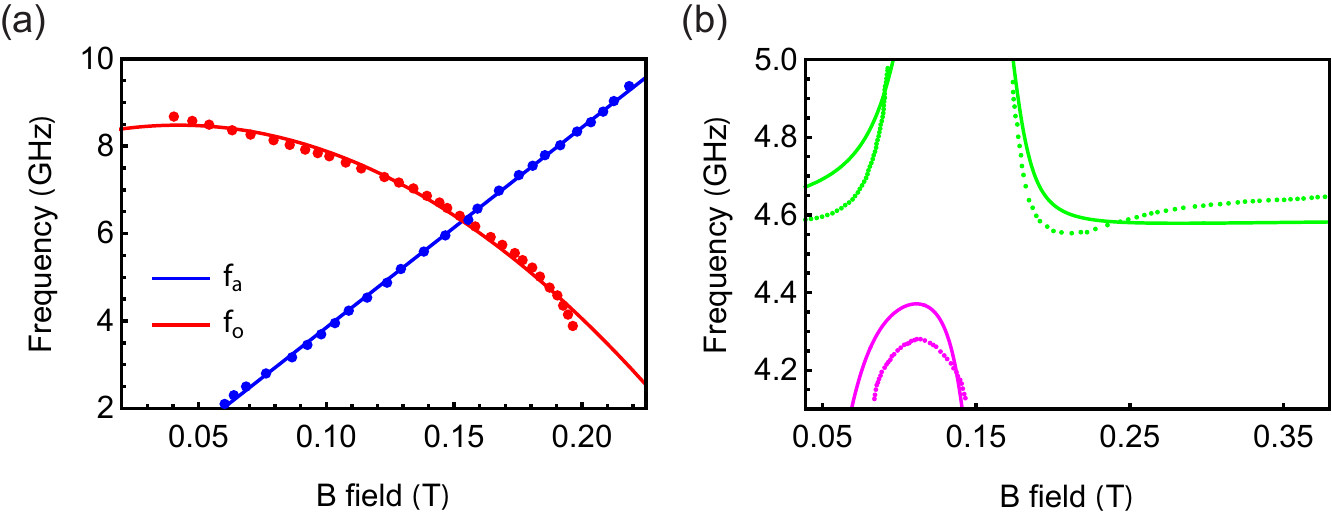}
\caption{\label{fig:fig10} (a) Polynomial fit (solid lines) to magnetic field dispersions of bare acoustic and optical AFMR modes for CrCl$_3$ crystal flake placed on copper transmission line. (b) Non-linear model fit to mode dispersion corresponding to magnon-photon coupling data of Fig.4(b) of the manuscript using magnetic field dependent functional form of the eigenmodes of three mode coupling Hamiltonian matrix (the dots represent the experimental data and the solid lines represent the fit).}
\end{figure*}
The reflection coefficient ($S_\textrm{11}$) for the system with two non-interacting magnon modes coupled with a cavity photon mode can be written according to input-output theory \cite{schuster_high-cooperativity_2010,abe_electron_2011,clerk_introduction_2010} as
\begin{equation} \label{eq2}
\begin{split}
S_{11}(f) & = 1-\frac{k_{e}}{i(f-f_0)+\frac{k_i+k_e}{2}+\frac{\lvert g_1/(2\pi) \rvert^2}{i(f-f_1)+\frac{\gamma_1/(2\pi)}{2}}+\frac{\lvert g_2/(2\pi) \rvert^2}{i(f-f_2)+\frac{\gamma_2/(2\pi)}{2}}}
\end{split}
\end{equation}
From Eq.\ref{eq2} we can obtain the functional form for $S_\textrm{11} (B,f)$ in case of CrCl$_3$ placed at current antinode of an NbN resonator, by inserting the dispersion relation of each mode with magnetic field. As the B field for this measurement is parallel to the plane of the SCPW resonator, the dispersion of cavity resonance frequency ($f_0$) can be taken to be constant within the field range of interest for observing magnon-photon coupling. We take this to be $f_0=4.6\  \textrm{GHz}$. For $k_i$ and $k_e$, we use the corresponding zero field values for the cavity (at zero field, cavity and AFMR modes are off-resonance), $k_i=\frac{\kappa_i}{2\pi}=0.0064\ \textrm{GHz}$ and $k_e=\frac{\kappa_e}{2\pi}=0.0145\ \textrm{GHz}$. For the field dependent functional form of the acoustic ($f_1$) and optical ($f_2$) AFMR mode frequencies in CrCl$_3$, we fitted linear and quadratic polynomials respectively to their field dispersions obtained from the transmission spectroscopy measurement using copper transmission line. Also, for the linewidths of the AFMR modes, we used $\frac{\gamma_1}{2\pi}=0.33\ \textrm{GHz}$ and $\frac{\gamma_2}{2\pi}=0.42\ \textrm{GHz}$, as their variation with magnetic field is small. \cite{kapoor_observation_2021} This fitting is shown in Fig.~\ref{fig:fig10}(a). For estimation of the coupling strengths, we used a three mode coupling Hamiltonian matrix, whose elements are described in the manuscript. Using magnetic field dependent functional form of the eigenvalues of this matrix, we did a non-linear model fitting to the mode dispersions of Fig.4(b) of manuscript. This fitting gives $\frac{g_1}{2\pi}=0.57\ \textrm{GHz}$ and $\frac{g_2}{2\pi}=0.37\ \textrm{GHz}$. The fitting has been shown in Fig.~\ref{fig:fig10}(b). From Fig.4(b) of the  manuscript, we note that there is a slight downward shift in cavity resonance frequency with increasing field, in the range 0 - 0.038 T which could possibly be attributed to change in magnetic state of CrCl$_3$ as it undergoes spin-flop transition. \cite{narath_spin-wave_1965,mcguire_magnetic_2017,kuhlow_magnetic_1982} Beyond this transition, initially antiparallel spins get in canted orientation with higher fields and give rise to net magnetic moment. We fit the data between 0.038 T and 0.380 T. The cooperativity has been calculated using these values as mentioned in the manuscript. We note that all the features of the observed data are not captured by the model considered. For further understanding some aspects of the data, a microscopic modeling and more detailed analysis is required.

As the crystal covers the entire SCPW resonator, the effective volume of interaction between the magnons and the microwave photons is essentially determined by the mode volume of the magnetic component of the RF field associated with the resonator. This mode volume can be calculated using the formula $V_\textrm{SCPW}=\frac{E_\textrm{SCPW}}{\frac{1}{2\mu_0} B_\textrm{max}^2}$, where $E_\textrm{SCPW}$ is the energy stored in the magnetic field of the SCPW resonator and $B_\textrm{max}$ is the maximum RF magnetic field of the SCPW resonator. We performed COMSOL simulation of the resonator to find $E_\textrm{SCPW}$ and $B_\textrm{max}$, which gives a magnetic mode volume approximately 0.006 mm$^3$ for the SCPW resonator. Furthermore, using the lattice parameters\cite{mcguire_magnetic_2017} of CrCl$_3$ and the magnetic component of the vacuum RF field of the SCPW resonator\cite{huebl_high_2013,lambert_identification_2015}, we calculate the magnon-photon coupling strength per spin to be $g_0=7.93$ Hz and corresponding net coupling strength for the mode volume considered to be $g=g_0 \sqrt{N}=1.09$ GHz, where $N$ is the total number of spins in the considered volume. Better understanding of the difference between $\frac{g}{2\pi}$ that are experimentally obtained, 0.57 GHz and 0.37 GHz for the acoustic and the optical modes respectively, and the estimate of 1.09 GHz, requires microscopic magnetic modeling. Furthermore, the coupling strengths associated with the acoustic and the optical magnons depend on exact geometry and symmetry, and hence a more accurate calculation taking into account symmetry and microscopic details is required for proper estimation for the coupling strengths.


\begin{thebibliography}{50}%
\makeatletter
\providecommand \@ifxundefined [1]{%
 \@ifx{#1\undefined}
}%
\providecommand \@ifnum [1]{%
 \ifnum #1\expandafter \@firstoftwo
 \else \expandafter \@secondoftwo
 \fi
}%
\providecommand \@ifx [1]{%
 \ifx #1\expandafter \@firstoftwo
 \else \expandafter \@secondoftwo
 \fi
}%
\providecommand \natexlab [1]{#1}%
\providecommand \enquote  [1]{``#1''}%
\providecommand \bibnamefont  [1]{#1}%
\providecommand \bibfnamefont [1]{#1}%
\providecommand \citenamefont [1]{#1}%
\providecommand \href@noop [0]{\@secondoftwo}%
\providecommand \href [0]{\begingroup \@sanitize@url \@href}%
\providecommand \@href[1]{\@@startlink{#1}\@@href}%
\providecommand \@@href[1]{\endgroup#1\@@endlink}%
\providecommand \@sanitize@url [0]{\catcode `\\12\catcode `\$12\catcode
  `\&12\catcode `\#12\catcode `\^12\catcode `\_12\catcode `\%12\relax}%
\providecommand \@@startlink[1]{}%
\providecommand \@@endlink[0]{}%
\providecommand \url  [0]{\begingroup\@sanitize@url \@url }%
\providecommand \@url [1]{\endgroup\@href {#1}{\urlprefix }}%
\providecommand \urlprefix  [0]{URL }%
\providecommand \Eprint [0]{\href }%
\providecommand \doibase [0]{http://dx.doi.org/}%
\providecommand \selectlanguage [0]{\@gobble}%
\providecommand \bibinfo  [0]{\@secondoftwo}%
\providecommand \bibfield  [0]{\@secondoftwo}%
\providecommand \translation [1]{[#1]}%
\providecommand \BibitemOpen [0]{}%
\providecommand \bibitemStop [0]{}%
\providecommand \bibitemNoStop [0]{.\EOS\space}%
\providecommand \EOS [0]{\spacefactor3000\relax}%
\providecommand \BibitemShut  [1]{\csname bibitem#1\endcsname}%
\let\auto@bib@innerbib\@empty
%</preamble>
\bibitem [{\citenamefont {Day}\ \emph {et~al.}(2003)\citenamefont {Day},
  \citenamefont {LeDuc}, \citenamefont {Mazin}, \citenamefont {Vayonakis},\
  and\ \citenamefont {Zmuidzinas}}]{day_broadband_2003}%
  \BibitemOpen
  \bibfield  {author} {\bibinfo {author} {\bibfnamefont {P.~K.}\ \bibnamefont
  {Day}}, \bibinfo {author} {\bibfnamefont {H.~G.}\ \bibnamefont {LeDuc}},
  \bibinfo {author} {\bibfnamefont {B.~A.}\ \bibnamefont {Mazin}}, \bibinfo
  {author} {\bibfnamefont {A.}~\bibnamefont {Vayonakis}}, \ and\ \bibinfo
  {author} {\bibfnamefont {J.}~\bibnamefont {Zmuidzinas}},\ }\href {\doibase
  10.1038/nature02037} {\bibfield  {journal} {\bibinfo  {journal} {Nature}\
  }\textbf {\bibinfo {volume} {425}},\ \bibinfo {pages} {817} (\bibinfo {year}
  {2003})}\BibitemShut {NoStop}%
\bibitem [{\citenamefont {Tholén}\ \emph {et~al.}(2007)\citenamefont
  {Tholén}, \citenamefont {Ergül}, \citenamefont {Doherty}, \citenamefont
  {Weber}, \citenamefont {Grégis},\ and\ \citenamefont
  {Haviland}}]{tholen_nonlinearities_2007}%
  \BibitemOpen
  \bibfield  {author} {\bibinfo {author} {\bibfnamefont {E.~A.}\ \bibnamefont
  {Tholén}}, \bibinfo {author} {\bibfnamefont {A.}~\bibnamefont {Ergül}},
  \bibinfo {author} {\bibfnamefont {E.~M.}\ \bibnamefont {Doherty}}, \bibinfo
  {author} {\bibfnamefont {F.~M.}\ \bibnamefont {Weber}}, \bibinfo {author}
  {\bibfnamefont {F.}~\bibnamefont {Grégis}}, \ and\ \bibinfo {author}
  {\bibfnamefont {D.~B.}\ \bibnamefont {Haviland}},\ }\href {\doibase
  10.1063/1.2750520} {\bibfield  {journal} {\bibinfo  {journal} {Applied
  Physics Letters}\ }\textbf {\bibinfo {volume} {90}},\ \bibinfo {pages}
  {253509} (\bibinfo {year} {2007})}\BibitemShut {NoStop}%
\bibitem [{\citenamefont {Castellanos-Beltran}\ and\ \citenamefont
  {Lehnert}(2007)}]{castellanos-beltran_widely_2007}%
  \BibitemOpen
  \bibfield  {author} {\bibinfo {author} {\bibfnamefont {M.~A.}\ \bibnamefont
  {Castellanos-Beltran}}\ and\ \bibinfo {author} {\bibfnamefont {K.~W.}\
  \bibnamefont {Lehnert}},\ }\href {\doibase 10.1063/1.2773988} {\bibfield
  {journal} {\bibinfo  {journal} {Applied Physics Letters}\ }\textbf {\bibinfo
  {volume} {91}},\ \bibinfo {pages} {083509} (\bibinfo {year}
  {2007})}\BibitemShut {NoStop}%
\bibitem [{\citenamefont {Wallraff}\ \emph {et~al.}(2004)\citenamefont
  {Wallraff}, \citenamefont {Schuster}, \citenamefont {Blais}, \citenamefont
  {Frunzio}, \citenamefont {Huang}, \citenamefont {Majer}, \citenamefont
  {Kumar}, \citenamefont {Girvin},\ and\ \citenamefont
  {Schoelkopf}}]{wallraff_strong_2004}%
  \BibitemOpen
  \bibfield  {author} {\bibinfo {author} {\bibfnamefont {A.}~\bibnamefont
  {Wallraff}}, \bibinfo {author} {\bibfnamefont {D.~I.}\ \bibnamefont
  {Schuster}}, \bibinfo {author} {\bibfnamefont {A.}~\bibnamefont {Blais}},
  \bibinfo {author} {\bibfnamefont {L.}~\bibnamefont {Frunzio}}, \bibinfo
  {author} {\bibfnamefont {R.-S.}\ \bibnamefont {Huang}}, \bibinfo {author}
  {\bibfnamefont {J.}~\bibnamefont {Majer}}, \bibinfo {author} {\bibfnamefont
  {S.}~\bibnamefont {Kumar}}, \bibinfo {author} {\bibfnamefont {S.~M.}\
  \bibnamefont {Girvin}}, \ and\ \bibinfo {author} {\bibfnamefont {R.~J.}\
  \bibnamefont {Schoelkopf}},\ }\href {\doibase 10.1038/nature02851} {\bibfield
   {journal} {\bibinfo  {journal} {Nature}\ }\textbf {\bibinfo {volume}
  {431}},\ \bibinfo {pages} {162} (\bibinfo {year} {2004})}\BibitemShut
  {NoStop}%
\bibitem [{\citenamefont {Regal}\ \emph {et~al.}(2008)\citenamefont {Regal},
  \citenamefont {Teufel},\ and\ \citenamefont
  {Lehnert}}]{regal_measuring_2008}%
  \BibitemOpen
  \bibfield  {author} {\bibinfo {author} {\bibfnamefont {C.~A.}\ \bibnamefont
  {Regal}}, \bibinfo {author} {\bibfnamefont {J.~D.}\ \bibnamefont {Teufel}}, \
  and\ \bibinfo {author} {\bibfnamefont {K.~W.}\ \bibnamefont {Lehnert}},\
  }\href {\doibase 10.1038/nphys974} {\bibfield  {journal} {\bibinfo  {journal}
  {Nature Physics}\ }\textbf {\bibinfo {volume} {4}},\ \bibinfo {pages} {555}
  (\bibinfo {year} {2008})}\BibitemShut {NoStop}%
\bibitem [{\citenamefont {Singh}\ \emph
  {et~al.}(2014{\natexlab{a}})\citenamefont {Singh}, \citenamefont {Bosman},
  \citenamefont {Schneider}, \citenamefont {Blanter}, \citenamefont
  {Castellanos-Gomez},\ and\ \citenamefont
  {Steele}}]{singh_optomechanical_2014}%
  \BibitemOpen
  \bibfield  {author} {\bibinfo {author} {\bibfnamefont {V.}~\bibnamefont
  {Singh}}, \bibinfo {author} {\bibfnamefont {S.~J.}\ \bibnamefont {Bosman}},
  \bibinfo {author} {\bibfnamefont {B.~H.}\ \bibnamefont {Schneider}}, \bibinfo
  {author} {\bibfnamefont {Y.~M.}\ \bibnamefont {Blanter}}, \bibinfo {author}
  {\bibfnamefont {A.}~\bibnamefont {Castellanos-Gomez}}, \ and\ \bibinfo
  {author} {\bibfnamefont {G.~A.}\ \bibnamefont {Steele}},\ }\href {\doibase
  10.1038/nnano.2014.168} {\bibfield  {journal} {\bibinfo  {journal} {Nature
  Nanotechnology}\ }\textbf {\bibinfo {volume} {9}},\ \bibinfo {pages} {820}
  (\bibinfo {year} {2014}{\natexlab{a}})}\BibitemShut {NoStop}%
\bibitem [{\citenamefont {Kubo}\ \emph {et~al.}(2010)\citenamefont {Kubo},
  \citenamefont {Ong}, \citenamefont {Bertet}, \citenamefont {Vion},
  \citenamefont {Jacques}, \citenamefont {Zheng}, \citenamefont {Dréau},
  \citenamefont {Roch}, \citenamefont {Auffeves}, \citenamefont {Jelezko},
  \citenamefont {Wrachtrup}, \citenamefont {Barthe}, \citenamefont {Bergonzo},\
  and\ \citenamefont {Esteve}}]{kubo_strong_2010}%
  \BibitemOpen
  \bibfield  {author} {\bibinfo {author} {\bibfnamefont {Y.}~\bibnamefont
  {Kubo}}, \bibinfo {author} {\bibfnamefont {F.~R.}\ \bibnamefont {Ong}},
  \bibinfo {author} {\bibfnamefont {P.}~\bibnamefont {Bertet}}, \bibinfo
  {author} {\bibfnamefont {D.}~\bibnamefont {Vion}}, \bibinfo {author}
  {\bibfnamefont {V.}~\bibnamefont {Jacques}}, \bibinfo {author} {\bibfnamefont
  {D.}~\bibnamefont {Zheng}}, \bibinfo {author} {\bibfnamefont
  {A.}~\bibnamefont {Dréau}}, \bibinfo {author} {\bibfnamefont {J.-F.}\
  \bibnamefont {Roch}}, \bibinfo {author} {\bibfnamefont {A.}~\bibnamefont
  {Auffeves}}, \bibinfo {author} {\bibfnamefont {F.}~\bibnamefont {Jelezko}},
  \bibinfo {author} {\bibfnamefont {J.}~\bibnamefont {Wrachtrup}}, \bibinfo
  {author} {\bibfnamefont {M.~F.}\ \bibnamefont {Barthe}}, \bibinfo {author}
  {\bibfnamefont {P.}~\bibnamefont {Bergonzo}}, \ and\ \bibinfo {author}
  {\bibfnamefont {D.}~\bibnamefont {Esteve}},\ }\href {\doibase
  10.1103/PhysRevLett.105.140502} {\bibfield  {journal} {\bibinfo  {journal}
  {Physical Review Letters}\ }\textbf {\bibinfo {volume} {105}},\ \bibinfo
  {pages} {140502} (\bibinfo {year} {2010})}\BibitemShut {NoStop}%
\bibitem [{\citenamefont {Amsüss}\ \emph {et~al.}(2011)\citenamefont
  {Amsüss}, \citenamefont {Koller}, \citenamefont {Nöbauer}, \citenamefont
  {Putz}, \citenamefont {Rotter}, \citenamefont {Sandner}, \citenamefont
  {Schneider}, \citenamefont {Schramböck}, \citenamefont {Steinhauser},
  \citenamefont {Ritsch}, \citenamefont {Schmiedmayer},\ and\ \citenamefont
  {Majer}}]{amsuss_cavity_2011}%
  \BibitemOpen
  \bibfield  {author} {\bibinfo {author} {\bibfnamefont {R.}~\bibnamefont
  {Amsüss}}, \bibinfo {author} {\bibfnamefont {C.}~\bibnamefont {Koller}},
  \bibinfo {author} {\bibfnamefont {T.}~\bibnamefont {Nöbauer}}, \bibinfo
  {author} {\bibfnamefont {S.}~\bibnamefont {Putz}}, \bibinfo {author}
  {\bibfnamefont {S.}~\bibnamefont {Rotter}}, \bibinfo {author} {\bibfnamefont
  {K.}~\bibnamefont {Sandner}}, \bibinfo {author} {\bibfnamefont
  {S.}~\bibnamefont {Schneider}}, \bibinfo {author} {\bibfnamefont
  {M.}~\bibnamefont {Schramböck}}, \bibinfo {author} {\bibfnamefont
  {G.}~\bibnamefont {Steinhauser}}, \bibinfo {author} {\bibfnamefont
  {H.}~\bibnamefont {Ritsch}}, \bibinfo {author} {\bibfnamefont
  {J.}~\bibnamefont {Schmiedmayer}}, \ and\ \bibinfo {author} {\bibfnamefont
  {J.}~\bibnamefont {Majer}},\ }\href {\doibase 10.1103/PhysRevLett.107.060502}
  {\bibfield  {journal} {\bibinfo  {journal} {Physical Review Letters}\
  }\textbf {\bibinfo {volume} {107}},\ \bibinfo {pages} {060502} (\bibinfo
  {year} {2011})}\BibitemShut {NoStop}%
\bibitem [{\citenamefont {Ranjan}\ \emph {et~al.}(2013)\citenamefont {Ranjan},
  \citenamefont {de~Lange}, \citenamefont {Schutjens}, \citenamefont
  {Debelhoir}, \citenamefont {Groen}, \citenamefont {Szombati}, \citenamefont
  {Thoen}, \citenamefont {Klapwijk}, \citenamefont {Hanson},\ and\
  \citenamefont {DiCarlo}}]{ranjan_probing_2013}%
  \BibitemOpen
  \bibfield  {author} {\bibinfo {author} {\bibfnamefont {V.}~\bibnamefont
  {Ranjan}}, \bibinfo {author} {\bibfnamefont {G.}~\bibnamefont {de~Lange}},
  \bibinfo {author} {\bibfnamefont {R.}~\bibnamefont {Schutjens}}, \bibinfo
  {author} {\bibfnamefont {T.}~\bibnamefont {Debelhoir}}, \bibinfo {author}
  {\bibfnamefont {J.~P.}\ \bibnamefont {Groen}}, \bibinfo {author}
  {\bibfnamefont {D.}~\bibnamefont {Szombati}}, \bibinfo {author}
  {\bibfnamefont {D.~J.}\ \bibnamefont {Thoen}}, \bibinfo {author}
  {\bibfnamefont {T.~M.}\ \bibnamefont {Klapwijk}}, \bibinfo {author}
  {\bibfnamefont {R.}~\bibnamefont {Hanson}}, \ and\ \bibinfo {author}
  {\bibfnamefont {L.}~\bibnamefont {DiCarlo}},\ }\href {\doibase
  10.1103/PhysRevLett.110.067004} {\bibfield  {journal} {\bibinfo  {journal}
  {Physical Review Letters}\ }\textbf {\bibinfo {volume} {110}},\ \bibinfo
  {pages} {067004} (\bibinfo {year} {2013})}\BibitemShut {NoStop}%
\bibitem [{\citenamefont {Tkalčec}\ \emph {et~al.}(2014)\citenamefont
  {Tkalčec}, \citenamefont {Probst}, \citenamefont {Rieger}, \citenamefont
  {Rotzinger}, \citenamefont {Wünsch}, \citenamefont {Kukharchyk},
  \citenamefont {Wieck}, \citenamefont {Siegel}, \citenamefont {Ustinov},\ and\
  \citenamefont {Bushev}}]{tkalcec_strong_2014}%
  \BibitemOpen
  \bibfield  {author} {\bibinfo {author} {\bibfnamefont {A.}~\bibnamefont
  {Tkalčec}}, \bibinfo {author} {\bibfnamefont {S.}~\bibnamefont {Probst}},
  \bibinfo {author} {\bibfnamefont {D.}~\bibnamefont {Rieger}}, \bibinfo
  {author} {\bibfnamefont {H.}~\bibnamefont {Rotzinger}}, \bibinfo {author}
  {\bibfnamefont {S.}~\bibnamefont {Wünsch}}, \bibinfo {author} {\bibfnamefont
  {N.}~\bibnamefont {Kukharchyk}}, \bibinfo {author} {\bibfnamefont {A.~D.}\
  \bibnamefont {Wieck}}, \bibinfo {author} {\bibfnamefont {M.}~\bibnamefont
  {Siegel}}, \bibinfo {author} {\bibfnamefont {A.~V.}\ \bibnamefont {Ustinov}},
  \ and\ \bibinfo {author} {\bibfnamefont {P.}~\bibnamefont {Bushev}},\ }\href
  {\doibase 10.1103/PhysRevB.90.075112} {\bibfield  {journal} {\bibinfo
  {journal} {Physical Review B}\ }\textbf {\bibinfo {volume} {90}},\ \bibinfo
  {pages} {075112} (\bibinfo {year} {2014})}\BibitemShut {NoStop}%
\bibitem [{\citenamefont {Zollitsch}\ \emph {et~al.}(2015)\citenamefont
  {Zollitsch}, \citenamefont {Mueller}, \citenamefont {Franke}, \citenamefont
  {Goennenwein}, \citenamefont {Brandt}, \citenamefont {Gross},\ and\
  \citenamefont {Huebl}}]{zollitsch_high_2015}%
  \BibitemOpen
  \bibfield  {author} {\bibinfo {author} {\bibfnamefont {C.~W.}\ \bibnamefont
  {Zollitsch}}, \bibinfo {author} {\bibfnamefont {K.}~\bibnamefont {Mueller}},
  \bibinfo {author} {\bibfnamefont {D.~P.}\ \bibnamefont {Franke}}, \bibinfo
  {author} {\bibfnamefont {S.~T.~B.}\ \bibnamefont {Goennenwein}}, \bibinfo
  {author} {\bibfnamefont {M.~S.}\ \bibnamefont {Brandt}}, \bibinfo {author}
  {\bibfnamefont {R.}~\bibnamefont {Gross}}, \ and\ \bibinfo {author}
  {\bibfnamefont {H.}~\bibnamefont {Huebl}},\ }\href {\doibase
  10.1063/1.4932658} {\bibfield  {journal} {\bibinfo  {journal} {Applied
  Physics Letters}\ }\textbf {\bibinfo {volume} {107}},\ \bibinfo {pages}
  {142105} (\bibinfo {year} {2015})}\BibitemShut {NoStop}%
\bibitem [{\citenamefont {Petta}\ \emph {et~al.}(2005)\citenamefont {Petta},
  \citenamefont {Johnson}, \citenamefont {Taylor}, \citenamefont {Laird},
  \citenamefont {Yacoby}, \citenamefont {Lukin}, \citenamefont {Marcus},
  \citenamefont {Hanson},\ and\ \citenamefont {Gossard}}]{petta_coherent_2005}%
  \BibitemOpen
  \bibfield  {author} {\bibinfo {author} {\bibfnamefont {J.~R.}\ \bibnamefont
  {Petta}}, \bibinfo {author} {\bibfnamefont {A.~C.}\ \bibnamefont {Johnson}},
  \bibinfo {author} {\bibfnamefont {J.~M.}\ \bibnamefont {Taylor}}, \bibinfo
  {author} {\bibfnamefont {E.~A.}\ \bibnamefont {Laird}}, \bibinfo {author}
  {\bibfnamefont {A.}~\bibnamefont {Yacoby}}, \bibinfo {author} {\bibfnamefont
  {M.~D.}\ \bibnamefont {Lukin}}, \bibinfo {author} {\bibfnamefont {C.~M.}\
  \bibnamefont {Marcus}}, \bibinfo {author} {\bibfnamefont {M.~P.}\
  \bibnamefont {Hanson}}, \ and\ \bibinfo {author} {\bibfnamefont {A.~C.}\
  \bibnamefont {Gossard}},\ }\href {\doibase 10.1126/science.1116955}
  {\bibfield  {journal} {\bibinfo  {journal} {Science}\ }\textbf {\bibinfo
  {volume} {309}},\ \bibinfo {pages} {2180} (\bibinfo {year}
  {2005})}\BibitemShut {NoStop}%
\bibitem [{\citenamefont {Nowack}\ \emph {et~al.}(2007)\citenamefont {Nowack},
  \citenamefont {Koppens}, \citenamefont {Nazarov},\ and\ \citenamefont
  {Vandersypen}}]{nowack_coherent_2007}%
  \BibitemOpen
  \bibfield  {author} {\bibinfo {author} {\bibfnamefont {K.~C.}\ \bibnamefont
  {Nowack}}, \bibinfo {author} {\bibfnamefont {F.~H.~L.}\ \bibnamefont
  {Koppens}}, \bibinfo {author} {\bibfnamefont {Y.~V.}\ \bibnamefont
  {Nazarov}}, \ and\ \bibinfo {author} {\bibfnamefont {L.~M.~K.}\ \bibnamefont
  {Vandersypen}},\ }\href {\doibase 10.1126/science.1148092} {\bibfield
  {journal} {\bibinfo  {journal} {Science}\ }\textbf {\bibinfo {volume}
  {318}},\ \bibinfo {pages} {1430} (\bibinfo {year} {2007})}\BibitemShut
  {NoStop}%
\bibitem [{\citenamefont {Schuster}\ \emph {et~al.}(2010)\citenamefont
  {Schuster}, \citenamefont {Sears}, \citenamefont {Ginossar}, \citenamefont
  {DiCarlo}, \citenamefont {Frunzio}, \citenamefont {Morton}, \citenamefont
  {Wu}, \citenamefont {Briggs}, \citenamefont {Buckley}, \citenamefont
  {Awschalom},\ and\ \citenamefont
  {Schoelkopf}}]{schuster_high-cooperativity_2010}%
  \BibitemOpen
  \bibfield  {author} {\bibinfo {author} {\bibfnamefont {D.~I.}\ \bibnamefont
  {Schuster}}, \bibinfo {author} {\bibfnamefont {A.~P.}\ \bibnamefont {Sears}},
  \bibinfo {author} {\bibfnamefont {E.}~\bibnamefont {Ginossar}}, \bibinfo
  {author} {\bibfnamefont {L.}~\bibnamefont {DiCarlo}}, \bibinfo {author}
  {\bibfnamefont {L.}~\bibnamefont {Frunzio}}, \bibinfo {author} {\bibfnamefont
  {J.~J.~L.}\ \bibnamefont {Morton}}, \bibinfo {author} {\bibfnamefont
  {H.}~\bibnamefont {Wu}}, \bibinfo {author} {\bibfnamefont {G.~A.~D.}\
  \bibnamefont {Briggs}}, \bibinfo {author} {\bibfnamefont {B.~B.}\
  \bibnamefont {Buckley}}, \bibinfo {author} {\bibfnamefont {D.~D.}\
  \bibnamefont {Awschalom}}, \ and\ \bibinfo {author} {\bibfnamefont {R.~J.}\
  \bibnamefont {Schoelkopf}},\ }\href {\doibase 10.1103/PhysRevLett.105.140501}
  {\bibfield  {journal} {\bibinfo  {journal} {Physical Review Letters}\
  }\textbf {\bibinfo {volume} {105}},\ \bibinfo {pages} {140501} (\bibinfo
  {year} {2010})}\BibitemShut {NoStop}%
\bibitem [{\citenamefont {Malissa}\ \emph {et~al.}(2013)\citenamefont
  {Malissa}, \citenamefont {Schuster}, \citenamefont {Tyryshkin}, \citenamefont
  {Houck},\ and\ \citenamefont {Lyon}}]{malissa_superconducting_2013}%
  \BibitemOpen
  \bibfield  {author} {\bibinfo {author} {\bibfnamefont {H.}~\bibnamefont
  {Malissa}}, \bibinfo {author} {\bibfnamefont {D.~I.}\ \bibnamefont
  {Schuster}}, \bibinfo {author} {\bibfnamefont {A.~M.}\ \bibnamefont
  {Tyryshkin}}, \bibinfo {author} {\bibfnamefont {A.~A.}\ \bibnamefont
  {Houck}}, \ and\ \bibinfo {author} {\bibfnamefont {S.~A.}\ \bibnamefont
  {Lyon}},\ }\href {\doibase 10.1063/1.4792205} {\bibfield  {journal} {\bibinfo
   {journal} {Review of Scientific Instruments}\ }\textbf {\bibinfo {volume}
  {84}},\ \bibinfo {pages} {025116} (\bibinfo {year} {2013})}\BibitemShut
  {NoStop}%
\bibitem [{\citenamefont {Benningshof}\ \emph {et~al.}(2013)\citenamefont
  {Benningshof}, \citenamefont {Mohebbi}, \citenamefont {Taminiau},
  \citenamefont {Miao},\ and\ \citenamefont
  {Cory}}]{benningshof_superconducting_2013}%
  \BibitemOpen
  \bibfield  {author} {\bibinfo {author} {\bibfnamefont {O.~W.~B.}\
  \bibnamefont {Benningshof}}, \bibinfo {author} {\bibfnamefont {H.~R.}\
  \bibnamefont {Mohebbi}}, \bibinfo {author} {\bibfnamefont {I.~A.~J.}\
  \bibnamefont {Taminiau}}, \bibinfo {author} {\bibfnamefont {G.~X.}\
  \bibnamefont {Miao}}, \ and\ \bibinfo {author} {\bibfnamefont {D.~G.}\
  \bibnamefont {Cory}},\ }\href {\doibase 10.1016/j.jmr.2013.01.010} {\bibfield
   {journal} {\bibinfo  {journal} {Journal of Magnetic Resonance}\ }\textbf
  {\bibinfo {volume} {230}},\ \bibinfo {pages} {84} (\bibinfo {year}
  {2013})}\BibitemShut {NoStop}%
\bibitem [{\citenamefont {Schmidt}\ \emph {et~al.}(2018)\citenamefont
  {Schmidt}, \citenamefont {Jenkins}, \citenamefont {Watanabe}, \citenamefont
  {Taniguchi},\ and\ \citenamefont {Steele}}]{schmidt_ballistic_2018}%
  \BibitemOpen
  \bibfield  {author} {\bibinfo {author} {\bibfnamefont {F.~E.}\ \bibnamefont
  {Schmidt}}, \bibinfo {author} {\bibfnamefont {M.~D.}\ \bibnamefont
  {Jenkins}}, \bibinfo {author} {\bibfnamefont {K.}~\bibnamefont {Watanabe}},
  \bibinfo {author} {\bibfnamefont {T.}~\bibnamefont {Taniguchi}}, \ and\
  \bibinfo {author} {\bibfnamefont {G.~A.}\ \bibnamefont {Steele}},\ }\href
  {\doibase 10.1038/s41467-018-06595-2} {\bibfield  {journal} {\bibinfo
  {journal} {Nature Communications}\ }\textbf {\bibinfo {volume} {9}},\
  \bibinfo {pages} {1} (\bibinfo {year} {2018})}\BibitemShut {NoStop}%
\bibitem [{\citenamefont {Kroll}\ \emph {et~al.}(2018)\citenamefont {Kroll},
  \citenamefont {Uilhoorn}, \citenamefont {Enden}, \citenamefont {Jong},
  \citenamefont {Watanabe}, \citenamefont {Taniguchi}, \citenamefont {Goswami},
  \citenamefont {Cassidy},\ and\ \citenamefont
  {Kouwenhoven}}]{kroll_magnetic_2018}%
  \BibitemOpen
  \bibfield  {author} {\bibinfo {author} {\bibfnamefont {J.~G.}\ \bibnamefont
  {Kroll}}, \bibinfo {author} {\bibfnamefont {W.}~\bibnamefont {Uilhoorn}},
  \bibinfo {author} {\bibfnamefont {K.~L. v.~d.}\ \bibnamefont {Enden}},
  \bibinfo {author} {\bibfnamefont {D.~d.}\ \bibnamefont {Jong}}, \bibinfo
  {author} {\bibfnamefont {K.}~\bibnamefont {Watanabe}}, \bibinfo {author}
  {\bibfnamefont {T.}~\bibnamefont {Taniguchi}}, \bibinfo {author}
  {\bibfnamefont {S.}~\bibnamefont {Goswami}}, \bibinfo {author} {\bibfnamefont
  {M.~C.}\ \bibnamefont {Cassidy}}, \ and\ \bibinfo {author} {\bibfnamefont
  {L.~P.}\ \bibnamefont {Kouwenhoven}},\ }\href {\doibase
  10.1038/s41467-018-07124-x} {\bibfield  {journal} {\bibinfo  {journal}
  {Nature Communications}\ }\textbf {\bibinfo {volume} {9}},\ \bibinfo {pages}
  {1} (\bibinfo {year} {2018})}\BibitemShut {NoStop}%
\bibitem [{\citenamefont {Wang}\ \emph {et~al.}(2019)\citenamefont {Wang},
  \citenamefont {Rodan-Legrain}, \citenamefont {Bretheau}, \citenamefont
  {Campbell}, \citenamefont {Kannan}, \citenamefont {Kim}, \citenamefont
  {Kjaergaard}, \citenamefont {Krantz}, \citenamefont {Samach}, \citenamefont
  {Yan}, \citenamefont {Yoder}, \citenamefont {Watanabe}, \citenamefont
  {Taniguchi}, \citenamefont {Orlando}, \citenamefont {Gustavsson},
  \citenamefont {Jarillo-Herrero},\ and\ \citenamefont
  {Oliver}}]{wang_coherent_2019}%
  \BibitemOpen
  \bibfield  {author} {\bibinfo {author} {\bibfnamefont {J.~I.-J.}\
  \bibnamefont {Wang}}, \bibinfo {author} {\bibfnamefont {D.}~\bibnamefont
  {Rodan-Legrain}}, \bibinfo {author} {\bibfnamefont {L.}~\bibnamefont
  {Bretheau}}, \bibinfo {author} {\bibfnamefont {D.~L.}\ \bibnamefont
  {Campbell}}, \bibinfo {author} {\bibfnamefont {B.}~\bibnamefont {Kannan}},
  \bibinfo {author} {\bibfnamefont {D.}~\bibnamefont {Kim}}, \bibinfo {author}
  {\bibfnamefont {M.}~\bibnamefont {Kjaergaard}}, \bibinfo {author}
  {\bibfnamefont {P.}~\bibnamefont {Krantz}}, \bibinfo {author} {\bibfnamefont
  {G.~O.}\ \bibnamefont {Samach}}, \bibinfo {author} {\bibfnamefont
  {F.}~\bibnamefont {Yan}}, \bibinfo {author} {\bibfnamefont {J.~L.}\
  \bibnamefont {Yoder}}, \bibinfo {author} {\bibfnamefont {K.}~\bibnamefont
  {Watanabe}}, \bibinfo {author} {\bibfnamefont {T.}~\bibnamefont {Taniguchi}},
  \bibinfo {author} {\bibfnamefont {T.~P.}\ \bibnamefont {Orlando}}, \bibinfo
  {author} {\bibfnamefont {S.}~\bibnamefont {Gustavsson}}, \bibinfo {author}
  {\bibfnamefont {P.}~\bibnamefont {Jarillo-Herrero}}, \ and\ \bibinfo {author}
  {\bibfnamefont {W.~D.}\ \bibnamefont {Oliver}},\ }\href {\doibase
  10.1038/s41565-018-0329-2} {\bibfield  {journal} {\bibinfo  {journal} {Nature
  Nanotechnology}\ }\textbf {\bibinfo {volume} {14}},\ \bibinfo {pages} {120}
  (\bibinfo {year} {2019})}\BibitemShut {NoStop}%
\bibitem [{\citenamefont {Gong}\ and\ \citenamefont
  {Zhang}(2019)}]{gong_two-dimensional_2019}%
  \BibitemOpen
  \bibfield  {author} {\bibinfo {author} {\bibfnamefont {C.}~\bibnamefont
  {Gong}}\ and\ \bibinfo {author} {\bibfnamefont {X.}~\bibnamefont {Zhang}},\
  }\href {\doibase 10.1126/science.aav4450} {\bibfield  {journal} {\bibinfo
  {journal} {Science}\ }\textbf {\bibinfo {volume} {363}},\ \bibinfo {pages}
  {eaav4450} (\bibinfo {year} {2019})}\BibitemShut {NoStop}%
\bibitem [{\citenamefont {Singh}\ \emph
  {et~al.}(2014{\natexlab{b}})\citenamefont {Singh}, \citenamefont {Schneider},
  \citenamefont {Bosman}, \citenamefont {Merkx},\ and\ \citenamefont
  {Steele}}]{singh_molybdenum-rhenium_2014}%
  \BibitemOpen
  \bibfield  {author} {\bibinfo {author} {\bibfnamefont {V.}~\bibnamefont
  {Singh}}, \bibinfo {author} {\bibfnamefont {B.~H.}\ \bibnamefont
  {Schneider}}, \bibinfo {author} {\bibfnamefont {S.~J.}\ \bibnamefont
  {Bosman}}, \bibinfo {author} {\bibfnamefont {E.~P.~J.}\ \bibnamefont
  {Merkx}}, \ and\ \bibinfo {author} {\bibfnamefont {G.~A.}\ \bibnamefont
  {Steele}},\ }\href {\doibase 10.1063/1.4903042} {\bibfield  {journal}
  {\bibinfo  {journal} {Applied Physics Letters}\ }\textbf {\bibinfo {volume}
  {105}},\ \bibinfo {pages} {222601} (\bibinfo {year}
  {2014}{\natexlab{b}})}\BibitemShut {NoStop}%
\bibitem [{\citenamefont {Calado}\ \emph {et~al.}(2015)\citenamefont {Calado},
  \citenamefont {Goswami}, \citenamefont {Nanda}, \citenamefont {Diez},
  \citenamefont {Akhmerov}, \citenamefont {Watanabe}, \citenamefont
  {Taniguchi}, \citenamefont {Klapwijk},\ and\ \citenamefont
  {Vandersypen}}]{calado_ballistic_2015}%
  \BibitemOpen
  \bibfield  {author} {\bibinfo {author} {\bibfnamefont {V.~E.}\ \bibnamefont
  {Calado}}, \bibinfo {author} {\bibfnamefont {S.}~\bibnamefont {Goswami}},
  \bibinfo {author} {\bibfnamefont {G.}~\bibnamefont {Nanda}}, \bibinfo
  {author} {\bibfnamefont {M.}~\bibnamefont {Diez}}, \bibinfo {author}
  {\bibfnamefont {A.~R.}\ \bibnamefont {Akhmerov}}, \bibinfo {author}
  {\bibfnamefont {K.}~\bibnamefont {Watanabe}}, \bibinfo {author}
  {\bibfnamefont {T.}~\bibnamefont {Taniguchi}}, \bibinfo {author}
  {\bibfnamefont {T.~M.}\ \bibnamefont {Klapwijk}}, \ and\ \bibinfo {author}
  {\bibfnamefont {L.~M.~K.}\ \bibnamefont {Vandersypen}},\ }\href {\doibase
  10.1038/nnano.2015.156} {\bibfield  {journal} {\bibinfo  {journal} {Nature
  Nanotechnology}\ }\textbf {\bibinfo {volume} {10}},\ \bibinfo {pages} {761}
  (\bibinfo {year} {2015})}\BibitemShut {NoStop}%
\bibitem [{\citenamefont {van Woerkom}\ \emph {et~al.}(2015)\citenamefont {van
  Woerkom}, \citenamefont {Geresdi},\ and\ \citenamefont
  {Kouwenhoven}}]{van_woerkom_one_2015}%
  \BibitemOpen
  \bibfield  {author} {\bibinfo {author} {\bibfnamefont {D.~J.}\ \bibnamefont
  {van Woerkom}}, \bibinfo {author} {\bibfnamefont {A.}~\bibnamefont
  {Geresdi}}, \ and\ \bibinfo {author} {\bibfnamefont {L.~P.}\ \bibnamefont
  {Kouwenhoven}},\ }\href {\doibase 10.1038/nphys3342} {\bibfield  {journal}
  {\bibinfo  {journal} {Nature Physics}\ }\textbf {\bibinfo {volume} {11}},\
  \bibinfo {pages} {547} (\bibinfo {year} {2015})}\BibitemShut {NoStop}%
\bibitem [{\citenamefont {Vissers}\ \emph {et~al.}(2010)\citenamefont
  {Vissers}, \citenamefont {Gao}, \citenamefont {Wisbey}, \citenamefont {Hite},
  \citenamefont {Tsuei}, \citenamefont {Corcoles}, \citenamefont {Steffen},\
  and\ \citenamefont {Pappas}}]{vissers_low_2010}%
  \BibitemOpen
  \bibfield  {author} {\bibinfo {author} {\bibfnamefont {M.~R.}\ \bibnamefont
  {Vissers}}, \bibinfo {author} {\bibfnamefont {J.}~\bibnamefont {Gao}},
  \bibinfo {author} {\bibfnamefont {D.~S.}\ \bibnamefont {Wisbey}}, \bibinfo
  {author} {\bibfnamefont {D.~A.}\ \bibnamefont {Hite}}, \bibinfo {author}
  {\bibfnamefont {C.~C.}\ \bibnamefont {Tsuei}}, \bibinfo {author}
  {\bibfnamefont {A.~D.}\ \bibnamefont {Corcoles}}, \bibinfo {author}
  {\bibfnamefont {M.}~\bibnamefont {Steffen}}, \ and\ \bibinfo {author}
  {\bibfnamefont {D.~P.}\ \bibnamefont {Pappas}},\ }\href {\doibase
  10.1063/1.3517252} {\bibfield  {journal} {\bibinfo  {journal} {Applied
  Physics Letters}\ }\textbf {\bibinfo {volume} {97}},\ \bibinfo {pages}
  {232509} (\bibinfo {year} {2010})}\BibitemShut {NoStop}%
\bibitem [{\citenamefont {Carter}\ \emph {et~al.}(2019)\citenamefont {Carter},
  \citenamefont {Khaire}, \citenamefont {Chang},\ and\ \citenamefont
  {Novosad}}]{carter_low-loss_2019}%
  \BibitemOpen
  \bibfield  {author} {\bibinfo {author} {\bibfnamefont {F.~W.}\ \bibnamefont
  {Carter}}, \bibinfo {author} {\bibfnamefont {T.}~\bibnamefont {Khaire}},
  \bibinfo {author} {\bibfnamefont {C.}~\bibnamefont {Chang}}, \ and\ \bibinfo
  {author} {\bibfnamefont {V.}~\bibnamefont {Novosad}},\ }\href {\doibase
  10.1063/1.5115276} {\bibfield  {journal} {\bibinfo  {journal} {Applied
  Physics Letters}\ }\textbf {\bibinfo {volume} {115}},\ \bibinfo {pages}
  {092602} (\bibinfo {year} {2019})}\BibitemShut {NoStop}%
\bibitem [{\citenamefont {Song}\ \emph
  {et~al.}(2009{\natexlab{a}})\citenamefont {Song}, \citenamefont {Heitmann},
  \citenamefont {DeFeo}, \citenamefont {Yu}, \citenamefont {McDermott},
  \citenamefont {Neeley}, \citenamefont {Martinis},\ and\ \citenamefont
  {Plourde}}]{song_microwave_2009}%
  \BibitemOpen
  \bibfield  {author} {\bibinfo {author} {\bibfnamefont {C.}~\bibnamefont
  {Song}}, \bibinfo {author} {\bibfnamefont {T.~W.}\ \bibnamefont {Heitmann}},
  \bibinfo {author} {\bibfnamefont {M.~P.}\ \bibnamefont {DeFeo}}, \bibinfo
  {author} {\bibfnamefont {K.}~\bibnamefont {Yu}}, \bibinfo {author}
  {\bibfnamefont {R.}~\bibnamefont {McDermott}}, \bibinfo {author}
  {\bibfnamefont {M.}~\bibnamefont {Neeley}}, \bibinfo {author} {\bibfnamefont
  {J.~M.}\ \bibnamefont {Martinis}}, \ and\ \bibinfo {author} {\bibfnamefont
  {B.~L.~T.}\ \bibnamefont {Plourde}},\ }\href {\doibase
  10.1103/PhysRevB.79.174512} {\bibfield  {journal} {\bibinfo  {journal}
  {Physical Review B}\ }\textbf {\bibinfo {volume} {79}},\ \bibinfo {pages}
  {174512} (\bibinfo {year} {2009}{\natexlab{a}})}\BibitemShut {NoStop}%
\bibitem [{\citenamefont {Ghirri}\ \emph {et~al.}(2015)\citenamefont {Ghirri},
  \citenamefont {Bonizzoni}, \citenamefont {Gerace}, \citenamefont {Sanna},
  \citenamefont {Cassinese},\ and\ \citenamefont
  {Affronte}}]{ghirri_yba2cu3o7_2015}%
  \BibitemOpen
  \bibfield  {author} {\bibinfo {author} {\bibfnamefont {A.}~\bibnamefont
  {Ghirri}}, \bibinfo {author} {\bibfnamefont {C.}~\bibnamefont {Bonizzoni}},
  \bibinfo {author} {\bibfnamefont {D.}~\bibnamefont {Gerace}}, \bibinfo
  {author} {\bibfnamefont {S.}~\bibnamefont {Sanna}}, \bibinfo {author}
  {\bibfnamefont {A.}~\bibnamefont {Cassinese}}, \ and\ \bibinfo {author}
  {\bibfnamefont {M.}~\bibnamefont {Affronte}},\ }\href {\doibase
  10.1063/1.4920930} {\bibfield  {journal} {\bibinfo  {journal} {Applied
  Physics Letters}\ }\textbf {\bibinfo {volume} {106}},\ \bibinfo {pages}
  {184101} (\bibinfo {year} {2015})}\BibitemShut {NoStop}%
\bibitem [{\citenamefont {Song}\ \emph
  {et~al.}(2009{\natexlab{b}})\citenamefont {Song}, \citenamefont {DeFeo},
  \citenamefont {Yu},\ and\ \citenamefont {Plourde}}]{song_reducing_2009}%
  \BibitemOpen
  \bibfield  {author} {\bibinfo {author} {\bibfnamefont {C.}~\bibnamefont
  {Song}}, \bibinfo {author} {\bibfnamefont {M.~P.}\ \bibnamefont {DeFeo}},
  \bibinfo {author} {\bibfnamefont {K.}~\bibnamefont {Yu}}, \ and\ \bibinfo
  {author} {\bibfnamefont {B.~L.~T.}\ \bibnamefont {Plourde}},\ }\href
  {\doibase 10.1063/1.3271523} {\bibfield  {journal} {\bibinfo  {journal}
  {Applied Physics Letters}\ }\textbf {\bibinfo {volume} {95}},\ \bibinfo
  {pages} {232501} (\bibinfo {year} {2009}{\natexlab{b}})}\BibitemShut
  {NoStop}%
\bibitem [{\citenamefont {Bothner}\ \emph {et~al.}(2011)\citenamefont
  {Bothner}, \citenamefont {Gaber}, \citenamefont {Kemmler}, \citenamefont
  {Koelle},\ and\ \citenamefont {Kleiner}}]{bothner_improving_2011}%
  \BibitemOpen
  \bibfield  {author} {\bibinfo {author} {\bibfnamefont {D.}~\bibnamefont
  {Bothner}}, \bibinfo {author} {\bibfnamefont {T.}~\bibnamefont {Gaber}},
  \bibinfo {author} {\bibfnamefont {M.}~\bibnamefont {Kemmler}}, \bibinfo
  {author} {\bibfnamefont {D.}~\bibnamefont {Koelle}}, \ and\ \bibinfo {author}
  {\bibfnamefont {R.}~\bibnamefont {Kleiner}},\ }\href {\doibase
  10.1063/1.3560480} {\bibfield  {journal} {\bibinfo  {journal} {Applied
  Physics Letters}\ }\textbf {\bibinfo {volume} {98}},\ \bibinfo {pages}
  {102504} (\bibinfo {year} {2011})}\BibitemShut {NoStop}%
\bibitem [{\citenamefont {Kroll}\ \emph {et~al.}(2019)\citenamefont {Kroll},
  \citenamefont {Borsoi}, \citenamefont {van~der Enden}, \citenamefont
  {Uilhoorn}, \citenamefont {de~Jong}, \citenamefont {Quintero-Pérez},
  \citenamefont {van Woerkom}, \citenamefont {Bruno}, \citenamefont {Plissard},
  \citenamefont {Car}, \citenamefont {Bakkers}, \citenamefont {Cassidy},\ and\
  \citenamefont {Kouwenhoven}}]{kroll_magnetic-field-resilient_2019}%
  \BibitemOpen
  \bibfield  {author} {\bibinfo {author} {\bibfnamefont {J.}~\bibnamefont
  {Kroll}}, \bibinfo {author} {\bibfnamefont {F.}~\bibnamefont {Borsoi}},
  \bibinfo {author} {\bibfnamefont {K.}~\bibnamefont {van~der Enden}}, \bibinfo
  {author} {\bibfnamefont {W.}~\bibnamefont {Uilhoorn}}, \bibinfo {author}
  {\bibfnamefont {D.}~\bibnamefont {de~Jong}}, \bibinfo {author} {\bibfnamefont
  {M.}~\bibnamefont {Quintero-Pérez}}, \bibinfo {author} {\bibfnamefont
  {D.}~\bibnamefont {van Woerkom}}, \bibinfo {author} {\bibfnamefont
  {A.}~\bibnamefont {Bruno}}, \bibinfo {author} {\bibfnamefont
  {S.}~\bibnamefont {Plissard}}, \bibinfo {author} {\bibfnamefont
  {D.}~\bibnamefont {Car}}, \bibinfo {author} {\bibfnamefont {E.}~\bibnamefont
  {Bakkers}}, \bibinfo {author} {\bibfnamefont {M.}~\bibnamefont {Cassidy}}, \
  and\ \bibinfo {author} {\bibfnamefont {L.}~\bibnamefont {Kouwenhoven}},\
  }\href {\doibase 10.1103/PhysRevApplied.11.064053} {\bibfield  {journal}
  {\bibinfo  {journal} {Physical Review Applied}\ }\textbf {\bibinfo {volume}
  {11}},\ \bibinfo {pages} {064053} (\bibinfo {year} {2019})}\BibitemShut
  {NoStop}%
\bibitem [{\citenamefont {Chockalingam}\ \emph {et~al.}(2008)\citenamefont
  {Chockalingam}, \citenamefont {Chand}, \citenamefont {Jesudasan},
  \citenamefont {Tripathi},\ and\ \citenamefont
  {Raychaudhuri}}]{chockalingam_superconducting_2008}%
  \BibitemOpen
  \bibfield  {author} {\bibinfo {author} {\bibfnamefont {S.~P.}\ \bibnamefont
  {Chockalingam}}, \bibinfo {author} {\bibfnamefont {M.}~\bibnamefont {Chand}},
  \bibinfo {author} {\bibfnamefont {J.}~\bibnamefont {Jesudasan}}, \bibinfo
  {author} {\bibfnamefont {V.}~\bibnamefont {Tripathi}}, \ and\ \bibinfo
  {author} {\bibfnamefont {P.}~\bibnamefont {Raychaudhuri}},\ }\href {\doibase
  10.1103/PhysRevB.77.214503} {\bibfield  {journal} {\bibinfo  {journal}
  {Physical Review B}\ }\textbf {\bibinfo {volume} {77}},\ \bibinfo {pages}
  {214503} (\bibinfo {year} {2008})}\BibitemShut {NoStop}%
\bibitem [{\citenamefont {Mondal}\ \emph {et~al.}(2011)\citenamefont {Mondal},
  \citenamefont {Kamlapure}, \citenamefont {Chand}, \citenamefont {Saraswat},
  \citenamefont {Kumar}, \citenamefont {Jesudasan}, \citenamefont {Benfatto},
  \citenamefont {Tripathi},\ and\ \citenamefont
  {Raychaudhuri}}]{mondal_phase_2011}%
  \BibitemOpen
  \bibfield  {author} {\bibinfo {author} {\bibfnamefont {M.}~\bibnamefont
  {Mondal}}, \bibinfo {author} {\bibfnamefont {A.}~\bibnamefont {Kamlapure}},
  \bibinfo {author} {\bibfnamefont {M.}~\bibnamefont {Chand}}, \bibinfo
  {author} {\bibfnamefont {G.}~\bibnamefont {Saraswat}}, \bibinfo {author}
  {\bibfnamefont {S.}~\bibnamefont {Kumar}}, \bibinfo {author} {\bibfnamefont
  {J.}~\bibnamefont {Jesudasan}}, \bibinfo {author} {\bibfnamefont
  {L.}~\bibnamefont {Benfatto}}, \bibinfo {author} {\bibfnamefont
  {V.}~\bibnamefont {Tripathi}}, \ and\ \bibinfo {author} {\bibfnamefont
  {P.}~\bibnamefont {Raychaudhuri}},\ }\href {\doibase
  10.1103/PhysRevLett.106.047001} {\bibfield  {journal} {\bibinfo  {journal}
  {Physical Review Letters}\ }\textbf {\bibinfo {volume} {106}},\ \bibinfo
  {pages} {047001} (\bibinfo {year} {2011})}\BibitemShut {NoStop}%
\bibitem [{\citenamefont {Kamlapure}\ \emph {et~al.}(2010)\citenamefont
  {Kamlapure}, \citenamefont {Mondal}, \citenamefont {Chand}, \citenamefont
  {Mishra}, \citenamefont {Jesudasan}, \citenamefont {Bagwe}, \citenamefont
  {Benfatto}, \citenamefont {Tripathi},\ and\ \citenamefont
  {Raychaudhuri}}]{kamlapure_measurement_2010}%
  \BibitemOpen
  \bibfield  {author} {\bibinfo {author} {\bibfnamefont {A.}~\bibnamefont
  {Kamlapure}}, \bibinfo {author} {\bibfnamefont {M.}~\bibnamefont {Mondal}},
  \bibinfo {author} {\bibfnamefont {M.}~\bibnamefont {Chand}}, \bibinfo
  {author} {\bibfnamefont {A.}~\bibnamefont {Mishra}}, \bibinfo {author}
  {\bibfnamefont {J.}~\bibnamefont {Jesudasan}}, \bibinfo {author}
  {\bibfnamefont {V.}~\bibnamefont {Bagwe}}, \bibinfo {author} {\bibfnamefont
  {L.}~\bibnamefont {Benfatto}}, \bibinfo {author} {\bibfnamefont
  {V.}~\bibnamefont {Tripathi}}, \ and\ \bibinfo {author} {\bibfnamefont
  {P.}~\bibnamefont {Raychaudhuri}},\ }\href {\doibase 10.1063/1.3314308}
  {\bibfield  {journal} {\bibinfo  {journal} {Applied Physics Letters}\
  }\textbf {\bibinfo {volume} {96}},\ \bibinfo {pages} {072509} (\bibinfo
  {year} {2010})}\BibitemShut {NoStop}%
\bibitem [{\citenamefont {MacNeill}\ \emph {et~al.}(2019)\citenamefont
  {MacNeill}, \citenamefont {Hou}, \citenamefont {Klein}, \citenamefont
  {Zhang}, \citenamefont {Jarillo-Herrero},\ and\ \citenamefont
  {Liu}}]{macneill_gigahertz_2019}%
  \BibitemOpen
  \bibfield  {author} {\bibinfo {author} {\bibfnamefont {D.}~\bibnamefont
  {MacNeill}}, \bibinfo {author} {\bibfnamefont {J.~T.}\ \bibnamefont {Hou}},
  \bibinfo {author} {\bibfnamefont {D.~R.}\ \bibnamefont {Klein}}, \bibinfo
  {author} {\bibfnamefont {P.}~\bibnamefont {Zhang}}, \bibinfo {author}
  {\bibfnamefont {P.}~\bibnamefont {Jarillo-Herrero}}, \ and\ \bibinfo {author}
  {\bibfnamefont {L.}~\bibnamefont {Liu}},\ }\href {\doibase
  10.1103/PhysRevLett.123.047204} {\bibfield  {journal} {\bibinfo  {journal}
  {Physical Review Letters}\ }\textbf {\bibinfo {volume} {123}},\ \bibinfo
  {pages} {047204} (\bibinfo {year} {2019})}\BibitemShut {NoStop}%
\bibitem [{\citenamefont {Zhang}\ \emph {et~al.}(2014)\citenamefont {Zhang},
  \citenamefont {Zou}, \citenamefont {Jiang},\ and\ \citenamefont
  {Tang}}]{zhang_strongly_2014}%
  \BibitemOpen
  \bibfield  {author} {\bibinfo {author} {\bibfnamefont {X.}~\bibnamefont
  {Zhang}}, \bibinfo {author} {\bibfnamefont {C.-L.}\ \bibnamefont {Zou}},
  \bibinfo {author} {\bibfnamefont {L.}~\bibnamefont {Jiang}}, \ and\ \bibinfo
  {author} {\bibfnamefont {H.~X.}\ \bibnamefont {Tang}},\ }\href {\doibase
  10.1103/PhysRevLett.113.156401} {\bibfield  {journal} {\bibinfo  {journal}
  {Physical Review Letters}\ }\textbf {\bibinfo {volume} {113}},\ \bibinfo
  {pages} {156401} (\bibinfo {year} {2014})}\BibitemShut {NoStop}%
\bibitem [{\citenamefont {Kapoor}\ \emph {et~al.}(2021)\citenamefont {Kapoor},
  \citenamefont {Mandal}, \citenamefont {Adak}, \citenamefont {Patankar},
  \citenamefont {Manni}, \citenamefont {Thamizhavel},\ and\ \citenamefont
  {Deshmukh}}]{kapoor_observation_2021}%
  \BibitemOpen
  \bibfield  {author} {\bibinfo {author} {\bibfnamefont {L.~N.}\ \bibnamefont
  {Kapoor}}, \bibinfo {author} {\bibfnamefont {S.}~\bibnamefont {Mandal}},
  \bibinfo {author} {\bibfnamefont {P.~C.}\ \bibnamefont {Adak}}, \bibinfo
  {author} {\bibfnamefont {M.}~\bibnamefont {Patankar}}, \bibinfo {author}
  {\bibfnamefont {S.}~\bibnamefont {Manni}}, \bibinfo {author} {\bibfnamefont
  {A.}~\bibnamefont {Thamizhavel}}, \ and\ \bibinfo {author} {\bibfnamefont
  {M.~M.}\ \bibnamefont {Deshmukh}},\ }\href {\doibase
  https://doi.org/10.1002/adma.202005105} {\bibfield  {journal} {\bibinfo
  {journal} {Advanced Materials}\ }\textbf {\bibinfo {volume} {33}},\ \bibinfo
  {pages} {2005105} (\bibinfo {year} {2021})}\BibitemShut {NoStop}%
\bibitem [{\citenamefont {Kim}\ \emph {et~al.}(2019)\citenamefont {Kim},
  \citenamefont {Yang}, \citenamefont {Li}, \citenamefont {Jiang},
  \citenamefont {Jin}, \citenamefont {Tao}, \citenamefont {Nichols},
  \citenamefont {Sfigakis}, \citenamefont {Zhong}, \citenamefont {Li},
  \citenamefont {Tian}, \citenamefont {Cory}, \citenamefont {Miao},
  \citenamefont {Shan}, \citenamefont {Mak}, \citenamefont {Lei}, \citenamefont
  {Sun}, \citenamefont {Zhao},\ and\ \citenamefont
  {Tsen}}]{kim_evolution_2019}%
  \BibitemOpen
  \bibfield  {author} {\bibinfo {author} {\bibfnamefont {H.~H.}\ \bibnamefont
  {Kim}}, \bibinfo {author} {\bibfnamefont {B.}~\bibnamefont {Yang}}, \bibinfo
  {author} {\bibfnamefont {S.}~\bibnamefont {Li}}, \bibinfo {author}
  {\bibfnamefont {S.}~\bibnamefont {Jiang}}, \bibinfo {author} {\bibfnamefont
  {C.}~\bibnamefont {Jin}}, \bibinfo {author} {\bibfnamefont {Z.}~\bibnamefont
  {Tao}}, \bibinfo {author} {\bibfnamefont {G.}~\bibnamefont {Nichols}},
  \bibinfo {author} {\bibfnamefont {F.}~\bibnamefont {Sfigakis}}, \bibinfo
  {author} {\bibfnamefont {S.}~\bibnamefont {Zhong}}, \bibinfo {author}
  {\bibfnamefont {C.}~\bibnamefont {Li}}, \bibinfo {author} {\bibfnamefont
  {S.}~\bibnamefont {Tian}}, \bibinfo {author} {\bibfnamefont {D.~G.}\
  \bibnamefont {Cory}}, \bibinfo {author} {\bibfnamefont {G.-X.}\ \bibnamefont
  {Miao}}, \bibinfo {author} {\bibfnamefont {J.}~\bibnamefont {Shan}}, \bibinfo
  {author} {\bibfnamefont {K.~F.}\ \bibnamefont {Mak}}, \bibinfo {author}
  {\bibfnamefont {H.}~\bibnamefont {Lei}}, \bibinfo {author} {\bibfnamefont
  {K.}~\bibnamefont {Sun}}, \bibinfo {author} {\bibfnamefont {L.}~\bibnamefont
  {Zhao}}, \ and\ \bibinfo {author} {\bibfnamefont {A.~W.}\ \bibnamefont
  {Tsen}},\ }\href {\doibase 10.1073/pnas.1902100116} {\bibfield  {journal}
  {\bibinfo  {journal} {Proceedings of the National Academy of Sciences}\
  }\textbf {\bibinfo {volume} {116}},\ \bibinfo {pages} {11131} (\bibinfo
  {year} {2019})}\BibitemShut {NoStop}%
\bibitem [{\citenamefont {Huebl}\ \emph {et~al.}(2013)\citenamefont {Huebl},
  \citenamefont {Zollitsch}, \citenamefont {Lotze}, \citenamefont {Hocke},
  \citenamefont {Greifenstein}, \citenamefont {Marx}, \citenamefont {Gross},\
  and\ \citenamefont {Goennenwein}}]{huebl_high_2013}%
  \BibitemOpen
  \bibfield  {author} {\bibinfo {author} {\bibfnamefont {H.}~\bibnamefont
  {Huebl}}, \bibinfo {author} {\bibfnamefont {C.~W.}\ \bibnamefont
  {Zollitsch}}, \bibinfo {author} {\bibfnamefont {J.}~\bibnamefont {Lotze}},
  \bibinfo {author} {\bibfnamefont {F.}~\bibnamefont {Hocke}}, \bibinfo
  {author} {\bibfnamefont {M.}~\bibnamefont {Greifenstein}}, \bibinfo {author}
  {\bibfnamefont {A.}~\bibnamefont {Marx}}, \bibinfo {author} {\bibfnamefont
  {R.}~\bibnamefont {Gross}}, \ and\ \bibinfo {author} {\bibfnamefont
  {S.~T.~B.}\ \bibnamefont {Goennenwein}},\ }\href {\doibase
  10.1103/PhysRevLett.111.127003} {\bibfield  {journal} {\bibinfo  {journal}
  {Physical Review Letters}\ }\textbf {\bibinfo {volume} {111}},\ \bibinfo
  {pages} {127003} (\bibinfo {year} {2013})}\BibitemShut {NoStop}%
\bibitem [{\citenamefont {Tabuchi}\ \emph {et~al.}(2014)\citenamefont
  {Tabuchi}, \citenamefont {Ishino}, \citenamefont {Ishikawa}, \citenamefont
  {Yamazaki}, \citenamefont {Usami},\ and\ \citenamefont
  {Nakamura}}]{tabuchi_hybridizing_2014}%
  \BibitemOpen
  \bibfield  {author} {\bibinfo {author} {\bibfnamefont {Y.}~\bibnamefont
  {Tabuchi}}, \bibinfo {author} {\bibfnamefont {S.}~\bibnamefont {Ishino}},
  \bibinfo {author} {\bibfnamefont {T.}~\bibnamefont {Ishikawa}}, \bibinfo
  {author} {\bibfnamefont {R.}~\bibnamefont {Yamazaki}}, \bibinfo {author}
  {\bibfnamefont {K.}~\bibnamefont {Usami}}, \ and\ \bibinfo {author}
  {\bibfnamefont {Y.}~\bibnamefont {Nakamura}},\ }\href {\doibase
  10.1103/PhysRevLett.113.083603} {\bibfield  {journal} {\bibinfo  {journal}
  {Physical Review Letters}\ }\textbf {\bibinfo {volume} {113}},\ \bibinfo
  {pages} {083603} (\bibinfo {year} {2014})}\BibitemShut {NoStop}%
\bibitem [{\citenamefont {Kwon}\ \emph {et~al.}(2018)\citenamefont {Kwon},
  \citenamefont {Fadavi~Roudsari}, \citenamefont {Benningshof}, \citenamefont
  {Tang}, \citenamefont {Mohebbi}, \citenamefont {Taminiau}, \citenamefont
  {Langenberg}, \citenamefont {Lee}, \citenamefont {Nichols}, \citenamefont
  {Cory},\ and\ \citenamefont {Miao}}]{kwon_magnetic_2018}%
  \BibitemOpen
  \bibfield  {author} {\bibinfo {author} {\bibfnamefont {S.}~\bibnamefont
  {Kwon}}, \bibinfo {author} {\bibfnamefont {A.}~\bibnamefont
  {Fadavi~Roudsari}}, \bibinfo {author} {\bibfnamefont {O.~W.~B.}\ \bibnamefont
  {Benningshof}}, \bibinfo {author} {\bibfnamefont {Y.-C.}\ \bibnamefont
  {Tang}}, \bibinfo {author} {\bibfnamefont {H.~R.}\ \bibnamefont {Mohebbi}},
  \bibinfo {author} {\bibfnamefont {I.~A.~J.}\ \bibnamefont {Taminiau}},
  \bibinfo {author} {\bibfnamefont {D.}~\bibnamefont {Langenberg}}, \bibinfo
  {author} {\bibfnamefont {S.}~\bibnamefont {Lee}}, \bibinfo {author}
  {\bibfnamefont {G.}~\bibnamefont {Nichols}}, \bibinfo {author} {\bibfnamefont
  {D.~G.}\ \bibnamefont {Cory}}, \ and\ \bibinfo {author} {\bibfnamefont
  {G.-X.}\ \bibnamefont {Miao}},\ }\href {\doibase 10.1063/1.5027003}
  {\bibfield  {journal} {\bibinfo  {journal} {Journal of Applied Physics}\
  }\textbf {\bibinfo {volume} {124}},\ \bibinfo {pages} {033903} (\bibinfo
  {year} {2018})}\BibitemShut {NoStop}%
\bibitem [{\citenamefont {Tinkham}(2004)}]{tinkham_introduction_2004}%
  \BibitemOpen
  \bibfield  {author} {\bibinfo {author} {\bibfnamefont {M.}~\bibnamefont
  {Tinkham}},\ }\href {https://books.google.co.in/books?id=VpUk3NfwDIkC} {\emph
  {\bibinfo {title} {Introduction to {Superconductivity}}}},\ Dover {Books} on
  {Physics} {Series}\ (\bibinfo  {publisher} {Dover Publications},\ \bibinfo
  {year} {2004})\BibitemShut {NoStop}%
\bibitem [{\citenamefont {Samkharadze}\ \emph {et~al.}(2016)\citenamefont
  {Samkharadze}, \citenamefont {Bruno}, \citenamefont {Scarlino}, \citenamefont
  {Zheng}, \citenamefont {DiVincenzo}, \citenamefont {DiCarlo},\ and\
  \citenamefont {Vandersypen}}]{samkharadze_high-kinetic-inductance_2016}%
  \BibitemOpen
  \bibfield  {author} {\bibinfo {author} {\bibfnamefont {N.}~\bibnamefont
  {Samkharadze}}, \bibinfo {author} {\bibfnamefont {A.}~\bibnamefont {Bruno}},
  \bibinfo {author} {\bibfnamefont {P.}~\bibnamefont {Scarlino}}, \bibinfo
  {author} {\bibfnamefont {G.}~\bibnamefont {Zheng}}, \bibinfo {author}
  {\bibfnamefont {D.~P.}\ \bibnamefont {DiVincenzo}}, \bibinfo {author}
  {\bibfnamefont {L.}~\bibnamefont {DiCarlo}}, \ and\ \bibinfo {author}
  {\bibfnamefont {L.~M.~K.}\ \bibnamefont {Vandersypen}},\ }\href {\doibase
  10.1103/PhysRevApplied.5.044004} {\bibfield  {journal} {\bibinfo  {journal}
  {Physical Review Applied}\ }\textbf {\bibinfo {volume} {5}},\ \bibinfo
  {pages} {044004} (\bibinfo {year} {2016})}\BibitemShut {NoStop}%
\bibitem [{\citenamefont {Mondal}\ \emph {et~al.}(2013)\citenamefont {Mondal},
  \citenamefont {Kamlapure}, \citenamefont {Ganguli}, \citenamefont
  {Jesudasan}, \citenamefont {Bagwe}, \citenamefont {Benfatto},\ and\
  \citenamefont {Raychaudhuri}}]{mondal_enhancement_2013}%
  \BibitemOpen
  \bibfield  {author} {\bibinfo {author} {\bibfnamefont {M.}~\bibnamefont
  {Mondal}}, \bibinfo {author} {\bibfnamefont {A.}~\bibnamefont {Kamlapure}},
  \bibinfo {author} {\bibfnamefont {S.~C.}\ \bibnamefont {Ganguli}}, \bibinfo
  {author} {\bibfnamefont {J.}~\bibnamefont {Jesudasan}}, \bibinfo {author}
  {\bibfnamefont {V.}~\bibnamefont {Bagwe}}, \bibinfo {author} {\bibfnamefont
  {L.}~\bibnamefont {Benfatto}}, \ and\ \bibinfo {author} {\bibfnamefont
  {P.}~\bibnamefont {Raychaudhuri}},\ }\href {\doibase 10.1038/srep01357}
  {\bibfield  {journal} {\bibinfo  {journal} {Scientific Reports}\ }\textbf
  {\bibinfo {volume} {3}},\ \bibinfo {pages} {1357} (\bibinfo {year}
  {2013})}\BibitemShut {NoStop}%
\bibitem [{\citenamefont {Maier-Flaig}\ \emph {et~al.}(2018)\citenamefont
  {Maier-Flaig}, \citenamefont {Goennenwein}, \citenamefont {Ohshima},
  \citenamefont {Shiraishi}, \citenamefont {Gross}, \citenamefont {Huebl},\
  and\ \citenamefont {Weiler}}]{maier-flaig_note_2018}%
  \BibitemOpen
  \bibfield  {author} {\bibinfo {author} {\bibfnamefont {H.}~\bibnamefont
  {Maier-Flaig}}, \bibinfo {author} {\bibfnamefont {S.~T.~B.}\ \bibnamefont
  {Goennenwein}}, \bibinfo {author} {\bibfnamefont {R.}~\bibnamefont
  {Ohshima}}, \bibinfo {author} {\bibfnamefont {M.}~\bibnamefont {Shiraishi}},
  \bibinfo {author} {\bibfnamefont {R.}~\bibnamefont {Gross}}, \bibinfo
  {author} {\bibfnamefont {H.}~\bibnamefont {Huebl}}, \ and\ \bibinfo {author}
  {\bibfnamefont {M.}~\bibnamefont {Weiler}},\ }\href {\doibase
  10.1063/1.5045135} {\bibfield  {journal} {\bibinfo  {journal} {Review of
  Scientific Instruments}\ }\textbf {\bibinfo {volume} {89}},\ \bibinfo {pages}
  {076101} (\bibinfo {year} {2018})}\BibitemShut {NoStop}%
\bibitem [{\citenamefont {Abe}\ \emph {et~al.}(2011)\citenamefont {Abe},
  \citenamefont {Wu}, \citenamefont {Ardavan},\ and\ \citenamefont
  {Morton}}]{abe_electron_2011}%
  \BibitemOpen
  \bibfield  {author} {\bibinfo {author} {\bibfnamefont {E.}~\bibnamefont
  {Abe}}, \bibinfo {author} {\bibfnamefont {H.}~\bibnamefont {Wu}}, \bibinfo
  {author} {\bibfnamefont {A.}~\bibnamefont {Ardavan}}, \ and\ \bibinfo
  {author} {\bibfnamefont {J.~J.~L.}\ \bibnamefont {Morton}},\ }\href {\doibase
  10.1063/1.3601930} {\bibfield  {journal} {\bibinfo  {journal} {Applied
  Physics Letters}\ }\textbf {\bibinfo {volume} {98}},\ \bibinfo {pages}
  {251108} (\bibinfo {year} {2011})},\ \bibinfo {note} {publisher: American
  Institute of Physics}\BibitemShut {NoStop}%
\bibitem [{\citenamefont {Clerk}\ \emph {et~al.}(2010)\citenamefont {Clerk},
  \citenamefont {Devoret}, \citenamefont {Girvin}, \citenamefont {Marquardt},\
  and\ \citenamefont {Schoelkopf}}]{clerk_introduction_2010}%
  \BibitemOpen
  \bibfield  {author} {\bibinfo {author} {\bibfnamefont {A.~A.}\ \bibnamefont
  {Clerk}}, \bibinfo {author} {\bibfnamefont {M.~H.}\ \bibnamefont {Devoret}},
  \bibinfo {author} {\bibfnamefont {S.~M.}\ \bibnamefont {Girvin}}, \bibinfo
  {author} {\bibfnamefont {F.}~\bibnamefont {Marquardt}}, \ and\ \bibinfo
  {author} {\bibfnamefont {R.~J.}\ \bibnamefont {Schoelkopf}},\ }\href
  {\doibase 10.1103/RevModPhys.82.1155} {\bibfield  {journal} {\bibinfo
  {journal} {Reviews of Modern Physics}\ }\textbf {\bibinfo {volume} {82}},\
  \bibinfo {pages} {1155} (\bibinfo {year} {2010})}\BibitemShut {NoStop}%
\bibitem [{\citenamefont {Lambert}\ \emph {et~al.}(2015)\citenamefont
  {Lambert}, \citenamefont {Haigh},\ and\ \citenamefont
  {Ferguson}}]{lambert_identification_2015}%
  \BibitemOpen
  \bibfield  {author} {\bibinfo {author} {\bibfnamefont {N.~J.}\ \bibnamefont
  {Lambert}}, \bibinfo {author} {\bibfnamefont {J.~A.}\ \bibnamefont {Haigh}},
  \ and\ \bibinfo {author} {\bibfnamefont {A.~J.}\ \bibnamefont {Ferguson}},\
  }\href {\doibase 10.1063/1.4907694} {\bibfield  {journal} {\bibinfo
  {journal} {Journal of Applied Physics}\ }\textbf {\bibinfo {volume} {117}},\
  \bibinfo {pages} {053910} (\bibinfo {year} {2015})}\BibitemShut {NoStop}%
\bibitem [{\citenamefont {Narath}\ and\ \citenamefont
  {Davis}(1965)}]{narath_spin-wave_1965}%
  \BibitemOpen
  \bibfield  {author} {\bibinfo {author} {\bibfnamefont {A.}~\bibnamefont
  {Narath}}\ and\ \bibinfo {author} {\bibfnamefont {H.~L.}\ \bibnamefont
  {Davis}},\ }\href {\doibase 10.1103/PhysRev.137.A163} {\bibfield  {journal}
  {\bibinfo  {journal} {Physical Review}\ }\textbf {\bibinfo {volume} {137}},\
  \bibinfo {pages} {A163} (\bibinfo {year} {1965})}\BibitemShut {NoStop}%
\bibitem [{\citenamefont {McGuire}\ \emph {et~al.}(2017)\citenamefont
  {McGuire}, \citenamefont {Clark}, \citenamefont {KC}, \citenamefont {Chance},
  \citenamefont {Jellison}, \citenamefont {Cooper}, \citenamefont {Xu},\ and\
  \citenamefont {Sales}}]{mcguire_magnetic_2017}%
  \BibitemOpen
  \bibfield  {author} {\bibinfo {author} {\bibfnamefont {M.~A.}\ \bibnamefont
  {McGuire}}, \bibinfo {author} {\bibfnamefont {G.}~\bibnamefont {Clark}},
  \bibinfo {author} {\bibfnamefont {S.}~\bibnamefont {KC}}, \bibinfo {author}
  {\bibfnamefont {W.~M.}\ \bibnamefont {Chance}}, \bibinfo {author}
  {\bibfnamefont {G.~E.}\ \bibnamefont {Jellison}}, \bibinfo {author}
  {\bibfnamefont {V.~R.}\ \bibnamefont {Cooper}}, \bibinfo {author}
  {\bibfnamefont {X.}~\bibnamefont {Xu}}, \ and\ \bibinfo {author}
  {\bibfnamefont {B.~C.}\ \bibnamefont {Sales}},\ }\href {\doibase
  10.1103/PhysRevMaterials.1.014001} {\bibfield  {journal} {\bibinfo  {journal}
  {Physical Review Materials}\ }\textbf {\bibinfo {volume} {1}},\ \bibinfo
  {pages} {014001} (\bibinfo {year} {2017})}\BibitemShut {NoStop}%
\bibitem [{\citenamefont {Kuhlow}(1982)}]{kuhlow_magnetic_1982}%
  \BibitemOpen
  \bibfield  {author} {\bibinfo {author} {\bibfnamefont {B.}~\bibnamefont
  {Kuhlow}},\ }\href {\doibase 10.1002/pssa.2210720116} {\bibfield  {journal}
  {\bibinfo  {journal} {physica status solidi (a)}\ }\textbf {\bibinfo {volume}
  {72}},\ \bibinfo {pages} {161} (\bibinfo {year} {1982})}\BibitemShut
  {NoStop}%
  \bibitem [{\citenamefont {Gao}\ \emph {et~al.}(2008)\citenamefont {Gao},
  \citenamefont {Daal}, \citenamefont {Vayonakis}, \citenamefont {Kumar},
  \citenamefont {Zmuidzinas}, \citenamefont {Sadoulet}, \citenamefont {Mazin},
  \citenamefont {Day},\ and\ \citenamefont {Leduc}}]{gao_experimental_2008}%
  \BibitemOpen
  \bibfield  {author} {\bibinfo {author} {\bibfnamefont {J.}~\bibnamefont
  {Gao}}, \bibinfo {author} {\bibfnamefont {M.}~\bibnamefont {Daal}}, \bibinfo
  {author} {\bibfnamefont {A.}~\bibnamefont {Vayonakis}}, \bibinfo {author}
  {\bibfnamefont {S.}~\bibnamefont {Kumar}}, \bibinfo {author} {\bibfnamefont
  {J.}~\bibnamefont {Zmuidzinas}}, \bibinfo {author} {\bibfnamefont
  {B.}~\bibnamefont {Sadoulet}}, \bibinfo {author} {\bibfnamefont {B.~A.}\
  \bibnamefont {Mazin}}, \bibinfo {author} {\bibfnamefont {P.~K.}\ \bibnamefont
  {Day}}, \ and\ \bibinfo {author} {\bibfnamefont {H.~G.}\ \bibnamefont
  {Leduc}},\ }\href {\doibase 10.1063/1.2906373} {\bibfield  {journal}
  {\bibinfo  {journal} {Applied Physics Letters}\ }\textbf {\bibinfo {volume}
  {92}},\ \bibinfo {pages} {152505} (\bibinfo {year} {2008})}\BibitemShut
  {NoStop}%
  \bibitem [{\citenamefont {Bothner}\ \emph {et~al.}(2012)\citenamefont
  {Bothner}, \citenamefont {Gaber}, \citenamefont {Kemmler}, \citenamefont
  {Koelle}, \citenamefont {Kleiner}, \citenamefont {Wünsch},\ and\
  \citenamefont {Siegel}}]{bothner_magnetic_2012}%
  \BibitemOpen
  \bibfield  {author} {\bibinfo {author} {\bibfnamefont {D.}~\bibnamefont
  {Bothner}}, \bibinfo {author} {\bibfnamefont {T.}~\bibnamefont {Gaber}},
  \bibinfo {author} {\bibfnamefont {M.}~\bibnamefont {Kemmler}}, \bibinfo
  {author} {\bibfnamefont {D.}~\bibnamefont {Koelle}}, \bibinfo {author}
  {\bibfnamefont {R.}~\bibnamefont {Kleiner}}, \bibinfo {author} {\bibfnamefont
  {S.}~\bibnamefont {Wünsch}}, \ and\ \bibinfo {author} {\bibfnamefont
  {M.}~\bibnamefont {Siegel}},\ }\href {\doibase 10.1103/PhysRevB.86.014517}
  {\bibfield  {journal} {\bibinfo  {journal} {Physical Review B}\ }\textbf
  {\bibinfo {volume} {86}},\ \bibinfo {pages} {014517} (\bibinfo {year}
  {2012})}\BibitemShut {NoStop}%
  \bibitem [{\citenamefont {Ji}\ \emph {et~al.}(1993)\citenamefont {Ji},
  \citenamefont {Rzchowski}, \citenamefont {Anand},\ and\ \citenamefont
  {Tinkham}}]{ji_magnetic-field-dependent_1993}%
  \BibitemOpen
  \bibfield  {author} {\bibinfo {author} {\bibfnamefont {L.}~\bibnamefont
  {Ji}}, \bibinfo {author} {\bibfnamefont {M.~S.}\ \bibnamefont {Rzchowski}},
  \bibinfo {author} {\bibfnamefont {N.}~\bibnamefont {Anand}}, \ and\ \bibinfo
  {author} {\bibfnamefont {M.}~\bibnamefont {Tinkham}},\ }\href {\doibase
  10.1103/PhysRevB.47.470} {\bibfield  {journal} {\bibinfo  {journal} {Physical
  Review B}\ }\textbf {\bibinfo {volume} {47}},\ \bibinfo {pages} {470}
  (\bibinfo {year} {1993})}\BibitemShut {NoStop}%
\bibitem [{\citenamefont
  {Raychaudhuri}(1996)}]{raychaudhuri_radio-frequency_1996}%
  \BibitemOpen
  \bibfield  {author} {\bibinfo {author} {\bibfnamefont {P.}~\bibnamefont
  {Raychaudhuri}},\ }\href {\doibase 10.1088/0953-2048/9/6/002} {\bibfield
  {journal} {\bibinfo  {journal} {Superconductor Science and Technology}\
  }\textbf {\bibinfo {volume} {9}},\ \bibinfo {pages} {447} (\bibinfo {year}
  {1996})}\BibitemShut {NoStop}%
\end{thebibliography}
\end{document}